\documentclass[journal,letter,10pt,twocolumn]{IEEEtran}
\pdfoutput=1
\usepackage{amsmath, amssymb, amscd, amsthm, amsfonts}
\usepackage{mathtools,lipsum,cuted}
\usepackage{enumerate}
\usepackage[pdftex]{graphicx}
\usepackage[caption=false,font=footnotesize,lofdepth,lotdepth]{subfig}
\usepackage{dblfloatfix}    
\usepackage{color}
\usepackage{algorithm}
\usepackage[noend]{algpseudocode}
\usepackage{comment}
\usepackage[noadjust]{cite}
\usepackage{bookmark}

\usepackage{tikz,ifthen}
\usetikzlibrary{positioning,arrows,backgrounds}
\usetikzlibrary{fit}
\usetikzlibrary{calc}

\newtheorem{theorem}{Theorem}
\newtheorem{lemma}{Lemma}
\newtheorem{corollary}{Corollary}
\newtheorem{proposition}{Proposition}
\newtheorem{definition}{Definition}

\usepackage{pgfplots}
\pgfplotsset{width=10cm,compat=1.9}

\usepackage{subfiles}
\usepackage[subtle,tracking = normal,charwidths = tight]{savetrees}

\usepackage{fancyhdr}

\usepackage[english]{babel}

\addto\captionsenglish{}

\usepackage{footnote}
\makesavenoteenv{tabular}
\makesavenoteenv{table}

\graphicspath{{../Figures/}}



\begin{document}

\title{On Minimizing Symbol Error Rate Over Fading \\ Channels with Low-Resolution Quantization}

\author{
   \noindent\IEEEauthorblockN{Neil Irwin Bernardo$^\dagger$, \textit{Graduate Student Member, IEEE}, Jingge Zhu, \textit{Member, IEEE},\\ and Jamie Evans, \textit{Senior Member, IEEE}
   }
  \thanks{Manuscript received Feb 16, 2021; revised June 21, 2021; accepted July 26, 2021. (\textit{Corresponding Author: Neil Irwin Bernardo})}
    \thanks{The work was supported in part by Australian Research Council under project DE210101497.}
    \thanks{$^\dagger$N.I. Bernardo acknowledges the Melbourne Research Scholarship of the University of Melbourne and the DOST-ERDT Faculty Development Fund of the Republic of the Philippines for sponsoring his doctoral studies.}
   \thanks{The authors are with the Department of Electrical and Electronics Engineering, The University of Melbourne, Parkville, VIC 3010, Australia. (e-mails: bernardon@student.unimelb.edu.au, jingge.zhu@unimelb.edu.au, jse@unimelb.edu.au).}
}

\maketitle
\begin{abstract}
We analyze the symbol error probability (SEP) of $M$-ary pulse amplitude modulation ($M$-PAM) receivers equipped with optimal low-resolution quantizers. We first show that the optimum detector can be reduced to a simple decision rule. Using this simplification, an exact SEP expression for quantized $M$-PAM receivers is obtained when Nakagami-$m$ fading channel is considered. The derived expression enables the optimization of the quantizer and/or constellation under the minimum SEP criterion. Our analysis of optimal quantization for equidistant $M$-PAM receiver reveals the existence of error floor which decays at a double exponential rate with increasing quantization bits, $b$. Moreover, by also allowing the transmitter to optimize the constellation based on the statistics of the fading channel, we prove that the error floor can be eliminated but at a lower decay exponent than the unquantized case. Characterization of this decay exponent is provided in this paper. We also expose the outage performance limitations of SEP-optimal uniform quantizers. To be more precise, its decay exponent does not improve with $b$. Lastly, we demonstrate that the decay exponent of a quantized receiver can be complemented by receive antenna diversity techniques.
\end{abstract}
\IEEEpeerreviewmaketitle
\section{Introduction}\label{section-intro}

\IEEEPARstart{H}{igh}-speed and high-resolution analog-to-digital converters (ADCs) are identified as one of the primary power consumers in a radio frequency (RF) receiver chain. Theoretical models for ADCs present power consumption as a quantity that scales exponentially with bit resolution and scales linearly with sampling rate \cite{Walden:1999}. Given this, one straightforward approach in solving the power consumption bottleneck of an RF chain is to simply reduce the sampling rate. However, the current trend in wireless research is geared towards the use of large bandwidths such as in millimeter wave (mmWave) cellular networks \cite{Rangan:2014} so using a low sampling rate would not be suitable for such systems. Thus, most studies on low-power receivers, such as \cite{Choi:2015,Choi:2016,Wen:2015,Wang:2019,Choi:2020}, have focused on designing practical detection strategies and analyzing the performance limits of receiver architectures with low-resolution ADCs.

A number of research results have shown that the use of low-precision ADCs in various communication systems offers substantial improvement in energy-efficiency and hardware cost while having little or negligible loss in achievable rate \cite{Liu:2019}. One of the earliest works on this topic has demonstrated that an $M$-PAM receiver with 2-3 quantization bits can already achieve 80-90\% of its unquantized channel capacity \cite{Singh:2009}. There also exists a rich body of literature \cite{Boccuzzi:2004,Sun:2010,Dabeer:2010,Liu:2018} showing that other receiver design functionalities (e.g. timing recovery, gain control, channel estimation) can be implemented in the low-resolution ADC regime with acceptable performance. The low-resolution ADC design approach is further justified by the hardware scaling laws observed in massive MIMO systems -- that is, for some fixed hardware quality, increasing the number of antennas reduces the impact of hardware distortion on the overall spectral efficiency (SE) performance \cite{Bjornson:2015}. In fact, the extreme case of using 1-bit ADCs in MIMO systems has been gaining significant research interest over the past few years due to its low-cost and scalable implementation \cite{Mezghani:2007,Risi:2014,Jacobsson:2015}. Aside from large-scale MIMO systems, communication with low-precision ADCs has also found a niche in energy-constrained applications such as wireless sensor networks and Internet-of-Things (IoT) \cite{Luo:2008,Gokceoglu:2017}.

Several information-theoretic studies \cite{Krone:2010, Koch:2013 ,Vu:2019} have established capacity limits of single-input single-output (SISO) channels with 1-bit ADC under various conditions. However, analysis of multi-bit quantization is much less tractable and analytical results are mostly based on simplified models. These models represent the inherent nonlinear characteristics of a quantizer as an additive noise. However, such models become inaccurate in the high SNR regime or when there are few antenna elements with coarsely quantized outputs \cite{Liu:2019}. Discrepancies between analytical and numerical results under these cases are mentioned in recent studies \cite{Singh:2009,Mezghani:2012,Orhan:2015,Azizzadeh:2019}. Traditional analytical models also assume either a mean square error (MSE)-optimal scalar quantizer \cite{Lloyd:1982} or a uniform quantizer. However, such design choices are not necessarily optimal if we intend to maximize the input-output mutual information or minimize the error rate of a communication link \cite{Liu:2019}. For example, a bit error rate (BER)-optimal ADC can have better error resiliency than a uniform or MSE-optimal quantizer even if the former has less quantization bits \cite{Narasimha:2012}. Studies on rate-optimal quantizer design also revealed that scalar quantization which maximizes information rate are not given by uniform and MSE-optimal quantizers \cite{Zeitler:2012}.

\begin{figure*}
    \centering
    \includegraphics[scale = .90,draft=false]{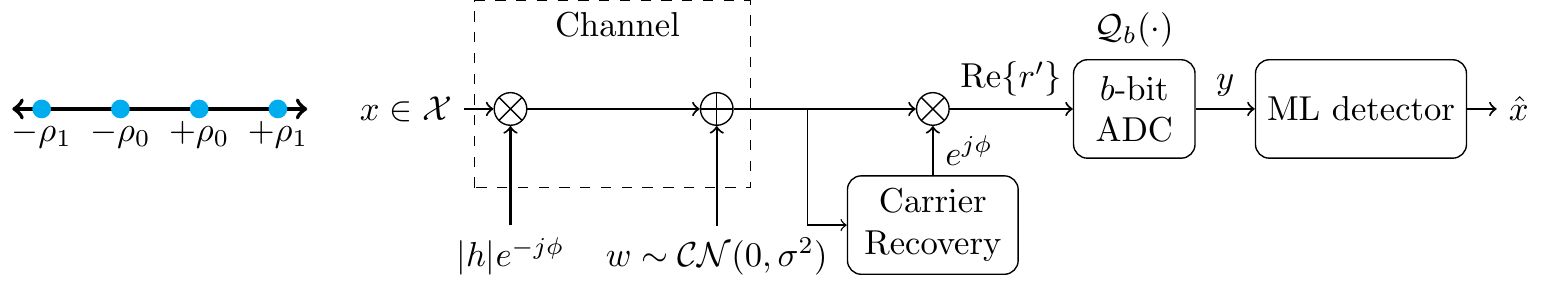}
     \caption{Coherent Detection of $M$-PAM Transmission over Fading Channel w/ Quantization}
    \label{fig:sys_model}
\end{figure*}

This paper investigates symbol error probability (SEP)-optimal quantization for communication over a general fading channel when the receiver is equipped with coarsely quantized ADC. In contrast to other works on communication systems equipped with low-resolution ADCs, the quantizer is designed based only on the statistics of the fading channel and does not need to adapt to small-scale fading variations. An essential aspect of our analysis is that we do not rely on simple additive noise models. Here, we deal with the nonlinear characteristics of the quantizer in order to gain some understanding on how the quantizer structure affects error performance. A simple SISO model is considered wherein the transmitter sends symbols drawn from an $M$-PAM constellation and a coherent receiver is equipped with a symmetric $b$-bit ADC. Error probability of an $M$-PAM receiver with finite ADC word length has been analyzed in \cite{Rizvi:2009} and \cite{Tuzlokov:2010}. However, the quantization process in these studies is modeled as an additive uniform noise and results are only applicable to uniform quantization. Uncoded error performance of quantized channels has been investigated previously for single-input multiple-output (SIMO) multiple access channel (MAC) \cite{Chowdhury:2015}. However, MSE-optimal quantizer is assumed and their result holds only for asymptotically large number of transmitters. Our motivation for investigating the coherent $M$-PAM case is to understand SEP-optimal quantization in fading channels in a single dimension and draw some insights that may aid in the analysis of more complex system models. Our main contributions are summarized as follows:
\begin{itemize}
    \item For $M$-PAM signaling with $M \geq 4$, we obtained an exact average SEP expression of the optimum maximum likelihood (ML) detector as a function of the quantizer structure and $M$-PAM amplitudes. The derived expression for the Nakagami-$m$ fading case is minimized numerically by optimizing the quantizer structure and/or the amplitudes of the $M$-PAM symbols. This result is presented in Theorem \ref{thm1}.
    \item For equidistant $M$-PAM signaling, we prove in Theorem \ref{thm2} that, with sufficient quantization bits, the lowest achievable SEP of the $M$-PAM receiver decreases exponentially with the shape parameter $m$ and double exponentially with quantization bits $b$. However, the SEP only goes down exponentially with increasing $b$ if the quantizer structure is restricted to uniform quantization (Theorem \ref{thm3}).
    \item By allowing the transmitter and receiver to jointly optimize the constellation and quantizer using statistical channel state information (CSI), we show that the error floor of a finite-resolution $M$-PAM receiver can be removed as long as $2^b > M - 2$ (Lemma \ref{lemma3}). We prove that the decay exponent (or diversity order\footnote{Diversity order is the asymptotic slope of the error probability as a function of SNR. Refer to equation (\ref{eq:DVO_definition}).}) of a low-resolution $M$-PAM receiver equipped with SEP-optimal quantizer is $m\frac{2^{b}-(M-2)}{2^{b}}$ when the $M$-PAM constellation is also optimized. This property, however, does not extend to uniform quantization. In fact, increasing the number of quantization bits does not improve the decay exponent of SEP-optimal uniform quantization. These results are presented in Theorem \ref{thm4} and Theorem \ref{thm5}.
\end{itemize}
The reader is referred to Table \ref{tab:summary_results} in Section \ref{section-conclusions} for a summary of these results. Finally, we give some insights in Section \ref{section-multiantenna} on how to extend the results to multiple antennas.

\section{System Model and ML Decision Rule for Quantized \\ Observations of $M$-PAM}\label{section-sysmodel-mldecision}

We consider a discrete-time SISO baseband channel model where a random complex-valued channel gain $h=|h|e^{-j\phi}$ is applied to the transmitted signal $x$ drawn from some $M$-PAM constellation set $\mathcal{X}$. As illustrated in Figure \ref{fig:sys_model}, the signal is corrupted by a circularly-symmetric complex-valued Gaussian noise $w \sim \mathcal{CN}\left(0,\sigma^2\right)$. Analog phase synchronization is then applied to the channel output by multiplying a $e^{j\phi}$ phase shift\footnote{The purpose of this is to analyze how quantization and fading jointly impact SEP in 1-D case. Notwithstanding, the use of Nakagami-$m$ in the analysis readily extends the results to when synchronization is absent. This is explained in Section \ref{section-SEP-MPAM}.}. This can be implemented in practice using analog phase-locked loop (PLL)-based circuit for rapid PAM carrier recovery such as Costas loop or a squaring loop \cite[198-199]{ling_proakis_2017}. The discrete-time baseband equivalent signal $r'$ can be expressed as
\begin{equation}\label{eq:sys_model_recv}
    r' = |h|x+w',\quad \text{where } w'=e^{j\phi}w\sim\mathcal{CN}(0,\sigma^2).
\end{equation}
\noindent The real part of $r'$ is then discretized by a $b$-bit ADC (i.e. $y = \mathcal{Q}_{b}\left(\mathrm{Re}\{|h|x+w'\}\right)$). In this work, we will only consider symmetric constellation sets and symmetric quantizers (i.e. $q_{-y} = -q_{y}$). The explicit quantization boundaries of the $b$-bit ADC are some real values $\{q_{\pm y}\}_{y = 1}^{y = 2^{b-1}-1}$ and implicit quantization boundaries are placed at $q_{0} = 0$ and $q_{\pm 2^{b-1}} = \pm \infty$. The ADC chooses $y = k$ whenever $\mathrm{Re}\{|h|x+w'\}$ falls inside the interval $\left(q_{k-1},q_{k}\right)$. We also assume that perfect CSI is available at the receiver. This assumption is justified by a previous work \cite{Dabeer:2010} which showed that accurate channel estimation is possible with low-precision ADC. Moreover, the receiver can use a high-precision ADC during channel estimation phase and then switch to a low-resolution ADC during data transmission phase. There will still be significant energy savings in this approach since the data transmission phase typically occupies a much larger portion of a coherence block \cite{Gayan:2020}. The likelihood of the ADC output $y$ given $|h|$ and $x$ is
\begin{equation} \label{eq:likelihood}
    \mathcal{L}\left(x\big||h|,y\right) = Q\left(\frac{q_{y-1} - |h|x}{\sqrt{\sigma^2/2}}\right) - 
         Q\left(\frac{q_{y} - |h|x}{\sqrt{\sigma^2/2}}\right), 
\end{equation}
where $Q(\cdot)$ is the tail probability of the standard Gaussian random variable. The ML detector for this quantized system chooses $\hat{x} = x^*$ if
\begin{equation}
    \label{eq:general_ML_rule}
    \hat{x}^{*} = \underset{x\in\mathcal{X}}{\arg \max}\;\mathcal{L}\left(x\big||h|,y\right).
\end{equation}
The detector, in its current form, is quite complex to use due to the Q-function terms. A simpler but equivalent detector is presented below which has no special functions or integration involved.
\begin{proposition}[\textbf{ML Detector for Finite-Resolution $M$-PAM Receiver}]\label{prop1}
Consider the model in Equation (\ref{eq:sys_model_recv}). For a given channel realization $h$ and ADC output $y$, $\hat{x}^*$, is
\begin{equation}
    \hat{x}^* = \underset{x\in \mathcal{X}}{\arg\min}\; \Big|\frac{q_{y}+q_{y-1}}{2} - |h|x\Big|.
\end{equation}
In other words, $|h|\hat{x}^*$ is closest to the midpoint of the quantization interval $\left(q_{y-1},q_{y}\right)$. Moreover, if $q_{y-1} = -\infty$ or $q_y = +\infty$, the midpoint is defined to be $-\infty$ and $+\infty$, respectively.
\end{proposition}

\begin{proof}
See Appendix \ref{proof_prop1}.
\end{proof}
We note that Proposition \ref{prop1} simplifies the scalar version of the ML detector given in \cite[Equation (13)]{Mezghani:2008} to a minimum distance detector. This proposition is used in the next section to derive the exact SEP of a finite-resolution $M$-PAM receiver. Analytical and simulation results are provided to illustrate the optimum SEP performance of low-resolution $M$-PAM receivers.

\section{Symbol Error Probability of $M$-PAM with $b$-bit\\ ADC}
\label{section-SEP-MPAM}

To gain insight on the error performance of a low-resolution $M$-PAM system, a simple case of $M$-PAM transmission (where $M \geq 4$ is a power of 2) and a coherent receiver with $b$-bit ADC is analyzed. We assume a real-valued constellation set $\mathcal{X}: \{\pm \rho_0,\pm\rho_1,\cdots,\pm\rho_{\frac{M}{2}-1}\}$. The SNR is the ratio of the average transmitted signal energy and additive noise and can be written as
\begin{equation}\label{eq:SNR_def}
    \begin{split}
        SNR = \frac{\mathbb{E}[x^2]}{\mathbb{E}[w'^2]} &= \frac{\mathcal{E}_{s}}{\sigma^2},\quad \text{where }\mathcal{E}_s = \frac{2\sum_{i=0}^{\frac{M}{2}-1}\rho_i^2}{M}.
    \end{split}
\end{equation}
Suppose the quantization boundaries of the $b$-bit ADC are $\{\pm q_{y}\}_{y=1}^{K}\cup\{0,\pm\infty\}$ where $K = 2^{b-1}-1$. We let $q_0 = 0$ and $\pm q_{K+1} = \pm \infty$. Since the constellation and quantizer are both symmetric, we can limit our analysis to the positive region. The ML rule is to pick a point closest to the midpoint of the quantization region. Let $\rho_{i-1}$, $\rho_i$, and $\rho_{i+1}$ be three positive adjacent PAM symbols in the constellation and $y > 0$ is the ADC output. We pick $\rho_i$ if 
\begin{equation*}
    \begin{split}
    \underbrace{\Big|\frac{q_{y-1}+q_{y}}{2} - |h|\rho_i\Big| < \Big|\frac{q_{y-1}+q_{y}}{2} - |h|\rho_{i-1}\Big|}_{\text{condition A}}
    \end{split}
\end{equation*}
and 
\begin{equation*}
    \begin{split}
        \underbrace{\Big|\frac{q_{y-1}+q_{y}}{2} - |h|\rho_i\Big| < \Big|\frac{q_{y-1}+q_{y}}{2} - |h|\rho_{i+1}\Big|}_{\text{condition B}}
    \end{split}
\end{equation*}
since $\rho_i$ is closer to the midpoint of $(q_{y-1},q_y)$ compared to its neighboring symbols. Only condition A (condition B) is needed to be satisfied $x = \rho_{\frac{M}{2}-1}$ ($x = \rho_{0}$). We can define decision regions $\mathcal{D}_{y,i}$, which chooses symbol $x = \rho_i$ if the ADC output is $y$, as
\begingroup
\allowdisplaybreaks
\begin{align}
    \mathcal{D}_{y,i}: &\quad     \mathcal{D}_{y,i}^{\text{L}} < |h|^2 < \mathcal{D}_{y,i}^{\text{U}},
\end{align}
\endgroup
such that when $y \in [1\mathrel{{.}\,{.}}\nobreak  K]$, we have
\begin{align*}
\mathcal{D}_{y,i}^{\text{L}} = \begin{cases}\left(\frac{q_{y-1}+q_y}{\rho_i+\rho_{i+1}}\right)^2,\;i\in[0\mathrel{{.}\,{.}}\nobreak \frac{M}{2}-2]\\
    \qquad0\qquad\;\;\;,\;i = \frac{M}{2}-1\end{cases}
\end{align*}
and
\begin{align*}
\mathcal{D}_{y,i}^{\text{U}} = \begin{cases}\left(\frac{q_{y-1}+q_y}{\rho_i+\rho_{i-1}}\right)^2,\;i\in[1\mathrel{{.}\,{.}}\nobreak \frac{M}{2}-1]\\
    \quad+\infty\qquad\;\;,\;i = 0\end{cases}.
\end{align*}
Furthermore, $\mathcal{D}_{K+1,i\in [0\mathrel{{.}\,{.}}\nobreak \frac{M}{2}-2]} = \emptyset$ and $\mathcal{D}_{K+1,\frac{M}{2}-1} = [0,+\infty)$. To obtain a general result about the SEP of the quantized receiver, we use a Nakagami-$m$ fading distribution for $|h|$. Suppose $|h| \sim \mathrm{Nakagami}(m,\Omega)$, then $Z = |h|^2 \sim \mathrm{Gamma}\left(m,\frac{\Omega}{m}\right)$ and has a probability distribution function (pdf)
\[f_Z(z) = \frac{m^m z^{m-1}e^{-\frac{mz}{\Omega}}}{\Gamma(m)\Omega^m}\quad\text{where $\Gamma(m) = \int_0^{\infty}t^{m-1}e^{-t^2}\;dt$}.
\]
$m$ and $\Omega$ are the shape and spread parameters of the distribution, respectively. We first prove a key lemma that simplifies the derivation of the SEP of a low-resolution $b$-bit  $M$-PAM receiver.
\begin{lemma}\label{lemma1}
Suppose we define $\mathcal{H}_{m,\Omega}(b,c,z_{\text{lo}},z_{\text{hi}})$ as the integral expression
\begin{align}\label{eq:H_m_omega}
    \mathcal{H}_{m,\Omega}\left(b,c,z_{\text{lo}},z_{\text{hi}}\right) = \int_{z_{lo}}^{z_{hi}}Q(-c+\sqrt{bz})f_Z(z)dz.
\end{align}
Then, $\mathcal{H}_{m,\Omega}\left(b,c,z_{\text{lo}},z_{\text{hi}}\right)$ has an exact finite series representation given by
\begingroup
\allowdisplaybreaks
\begin{align*}
    &\textbf{when $c > 0$}:\\
    &\quad Q(-c + \sqrt{bz_{\text{lo}}})\frac{\Gamma\left(m,\frac{m}{\Omega}z_{lo}\right)}{\Gamma(m)}- Q(-c + \sqrt{bz_{\text{hi}}})\frac{\Gamma\left(m,\frac{m}{\Omega}z_{hi}\right)}{\Gamma(m)}\\
    &-\sum_{r=0}^{m-1}\sum_{l=0}^{2r}\frac{\left( \frac{m}{\Omega b}\right)^r\binom{2r}{l}\exp\left(-\frac{c^2}{2}\left(\frac{\frac{2m}{\Omega b}}{\frac{2m}{\Omega b}+1}\right)\right) c^{2r-l}\mathcal{F}(u_{hi},u_{lo},l)}{\sqrt{2\pi}r!\left(\frac{2m+\Omega b}{\Omega b}\right)^{2r-0.5(l-1)}}\\
    &\textbf{when $c = 0$}:\\
    &\quad Q(\sqrt{bz_{\text{lo}}})\frac{\Gamma\left(m,\frac{m}{\Omega}z_{lo}\right)}{\Gamma(m)}- Q(\sqrt{bz_{\text{hi}}})\frac{\Gamma\left(m,\frac{m}{\Omega}z_{hi}\right)}{\Gamma(m)}\\
    &\quad\qquad\qquad-\sum_{r=0}^{m-1}\frac{\left(\frac{m}{\Omega b + 2m}\right)^r\sqrt{\frac{\Omega b}{\Omega b + 2m }}\mathcal{F}(u_{hi},u_{lo},2r)}{\sqrt{2\pi}r!}
\end{align*}
\endgroup
where
\begin{equation*}
    \begin{split}
         &u_{hi} = \frac{-c+\left(\frac{2m}{\Omega b} + 1\right)\sqrt{bz_{\text{hi}}}}{\sqrt{\frac{2m}{\Omega b}+1}}\;\;\text{ and }\;\; u_{lo} = \frac{-c+\left(\frac{2m}{\Omega b} + 1\right)\sqrt{bz_{\text{lo}}}}{\sqrt{\frac{2m}{\Omega b}+1}}\\
    \end{split}
\end{equation*}
for $m \in \mathbb{Z}$ . $\Gamma(m,x) = \int_x^{\infty}t^{m-1}e^{-t^2}\;dt$ is the upper incomplete Gamma function. The function $\mathcal{F}(a,b,l)$ is
\begin{equation}
\label{eq:F_function}
    \begin{split}
    \mathcal{F}(a,b,l) &= \int^a_{b} u^{l}e^{-\frac{u^2}{2}} \;du\\
    &=\begin{cases}-\text{sgn}(u)^{l+1}\left[\frac{\Gamma\left(\frac{l+1}{2},\frac{u^2}{2}\right)}{2^{\frac{1-l}{2}}}- \frac{\sqrt{\pi}(l-1)!!}{\sqrt{2}}\right]\bigg|_{b}^a,\text{$l$ is even}\\
    -\text{sgn}(u)^{l+1}\left[\frac{\Gamma\left(\frac{l+1}{2},\frac{u^2}{2}\right)}{2^{\frac{1-l}{2}}} \right]\bigg|_{b}^a,\text{$l$ is odd}
    \end{cases}
    \end{split}
\end{equation}
for $l\in\mathbb{Z}$ and $a,b\in\mathbb{R}$. $\text{sgn}(\cdot)$ and $(\cdot)!!$ are signum and double factorial functions, respectively.
\end{lemma}

\begin{proof}
See Appendix \ref{proof_lemma1}.
\end{proof}

\begin{figure}[t]
    \centering
    \includegraphics[scale = .6]{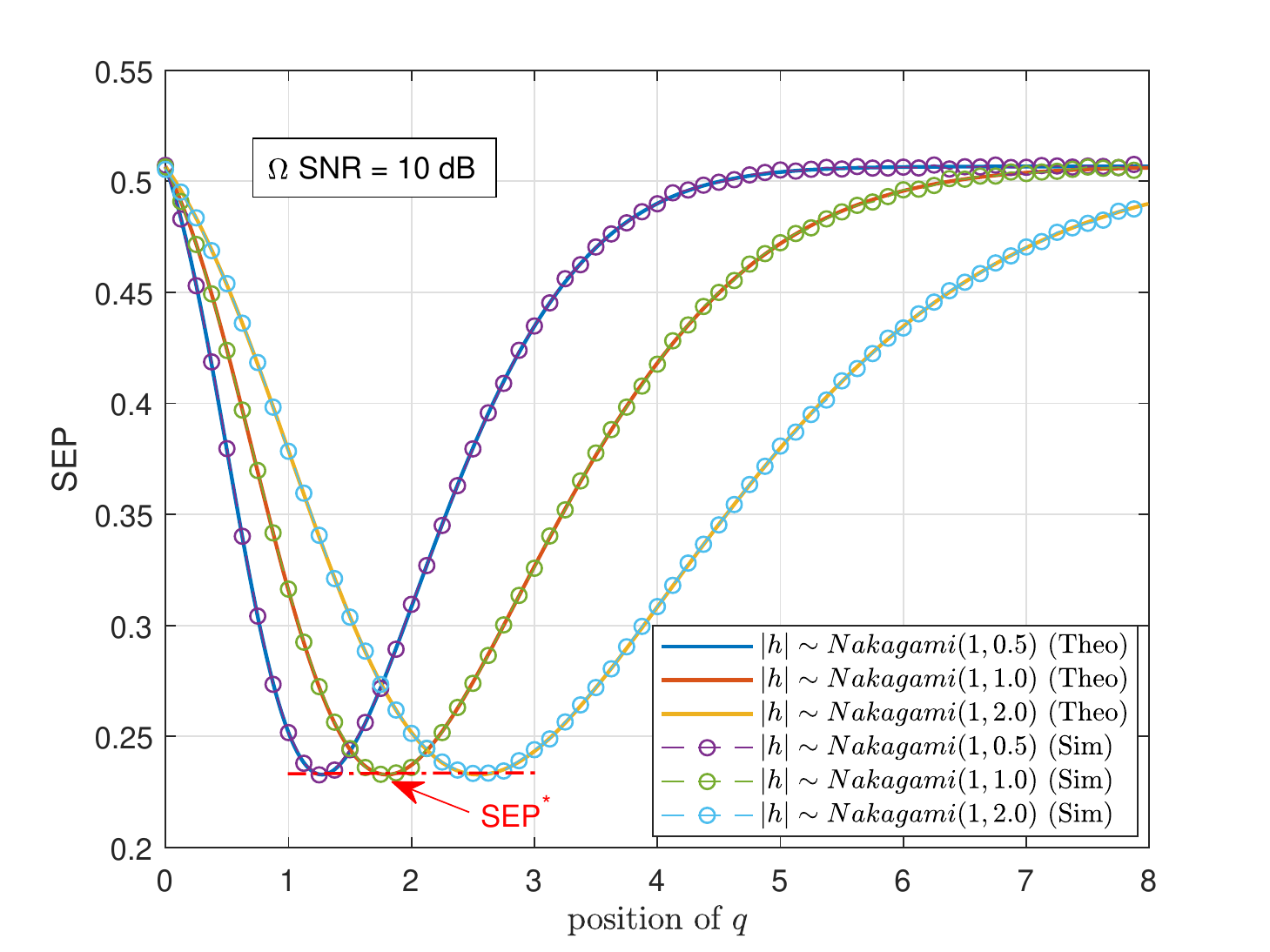}
    \caption{Symbol Error Probability vs. $q_1$ for $\Omega = \{0.5,1.0,2.0\}$ and $m=1$ (2-bit 4-PAM case) ($\Omega SNR$ = 10 dB)}
    \label{fig:SEP_vs_c}
\end{figure}

We now derive an exact analytical expression of the SEP in terms of $\mathcal{H}_{m,\Omega}(\cdot)$.
\begin{theorem}[\textbf{Average SEP of Finite-Resolution $M$-PAM Receiver}]\label{thm1}
Let $q_y$ be the $y$-th quantization boundary, $\mathcal{E}_s$ be the average symbol energy, and $\sigma^2$ be the noise variance. The average SEP of a low-resolution $b$-bit receiver architecture with $M$-PAM and Nakagami-$m$ fading is given by
\begin{align*}
     P_{\text{e}} =&  1 - \frac{2}{M}\sum_{y=1}^{2^{b-1}}\sum_{i=0}^{\frac{M}{2}-1}\Bigg\{\mathcal{H}_{m,\Omega}\left(SNR_i,\frac{\sqrt{2}q_{y}}{\sigma},\mathcal{D}_{y,i}^{\text{L}},\mathcal{D}_{y,i}^{\text{U}}\right) \\
     &\qquad\qquad\qquad- \mathcal{H}_{m,\Omega}\left(SNR_i,\frac{\sqrt{2}q_{y-1}}{\sigma},\mathcal{D}_{y,i}^{\text{L}},\mathcal{D}_{y,i}^{\text{U}}\right)\Bigg\},
\end{align*}
where $SNR_i = \frac{2\rho_i^2 SNR}{\mathcal{E}_s^2}$ and $m\in \mathbb{Z}$.
\end{theorem}

\begin{proof}
The general expression for the SEP is
\begingroup
\allowdisplaybreaks
\begin{align}\label{eq:gen_SEP_exp}
    P_e(SNR) =&  1 - \frac{2}{M}\sum_{y=1}^{2^{b-1}}\sum_{i=0}^{\frac{M}{2}-1}\underbrace{\int_{\mathcal{D}_{y,i}}f(y|z,x=\rho_i)f_Z(z)\;dz}_{\text{$\mathbb{P}$\{correctly detecting $x = \rho_i$\}}},
\end{align}
where
\begin{align*}
   f(y|z,x=\rho_i) =& Q\left(-\frac{\sqrt{2}q_{y}}{\sigma}+\frac{\sqrt{2z\rho_i^2 SNR}}{\mathcal{E}_{s}}\right)\\
   &\qquad-Q\left(-\frac{\sqrt{2}q_{y-1}}{\sigma}+\frac{\sqrt{2z\rho_i^2 SNR}}{\mathcal{E}_{s}}\right).
\end{align*}
\endgroup
The integral term in (\ref{eq:gen_SEP_exp}) has the same form as (\ref{eq:H_m_omega}). Thus, the average SEP expression can be simplified to an exact finite series by applying Lemma \ref{lemma1} on equation (\ref{eq:gen_SEP_exp}).
\end{proof}
While our exact expression in Theorem \ref{thm1} generalizes only to integer values of $m$, equation (\ref{eq:gen_SEP_exp}) can be used to numerically compute the SEP for any $m \geq \frac{1}{2}$. We also point out that Theorem \ref{thm1} can be applied to $M$-PAM receivers without analog carrier recovery prior to the ADC stage provided they know $\mathrm{Re}\{h\}$. The receiver can simply invert the numbering of the quantization regions whenever $\mathrm{Re}\{h\} < 0$. The analysis above still holds but $|h|$ is replaced with $|\mathrm{Re}\{h\}|$, which is still a Nakagami random variable with lower shape parameter \cite[eq. 23a and 23b]{Mallik:2010}. This, however, incurs performance degradation since the signal energy placed along the quadrature component due to channel phase rotation is thrown away.
\begin{figure}[t]
  \subfloat[]{
        \includegraphics[width=.5\textwidth,draft = false]{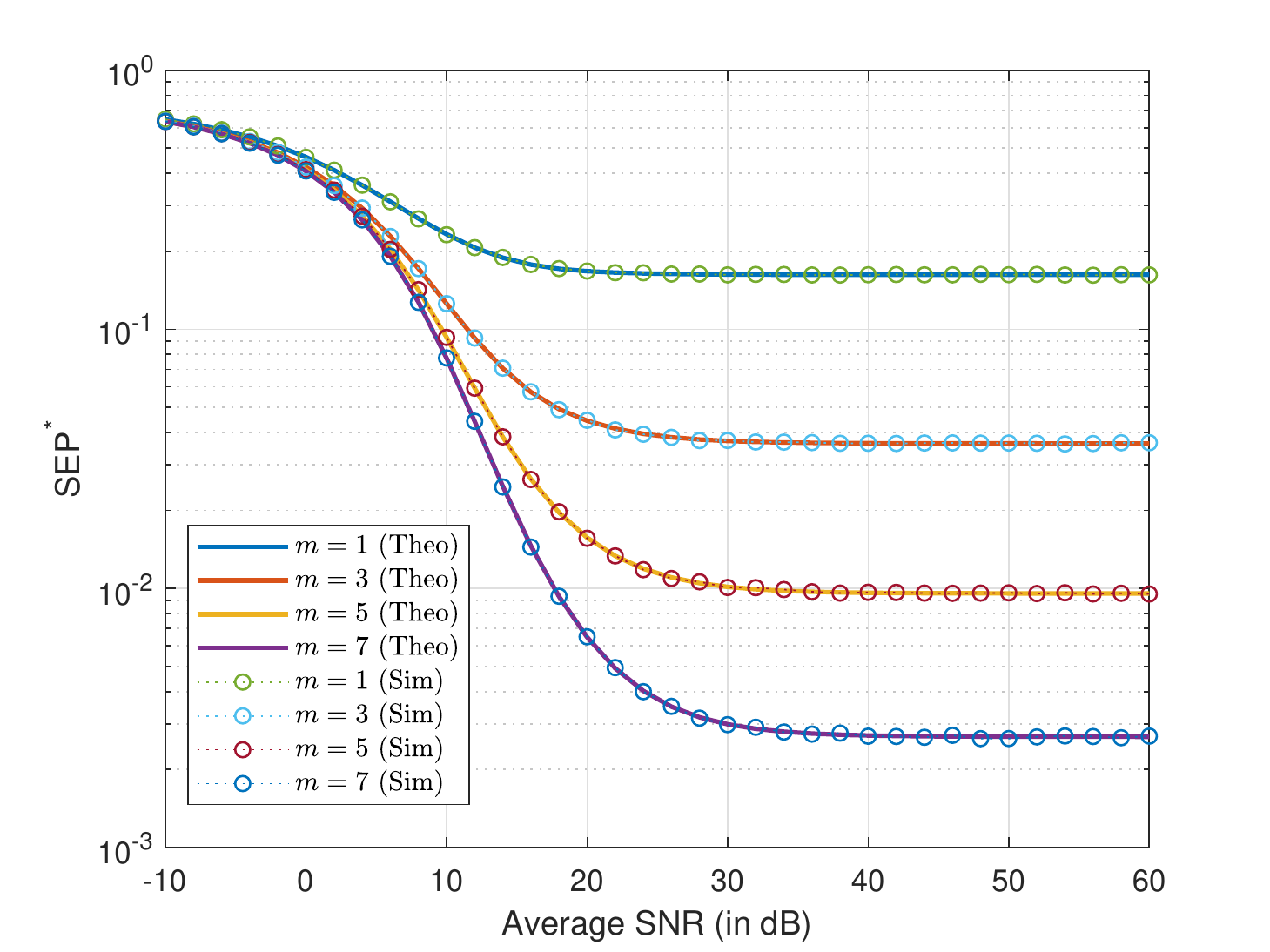}
    }
   \vskip0cm
    \subfloat[]{
        \includegraphics[width=.5\textwidth,draft=false]{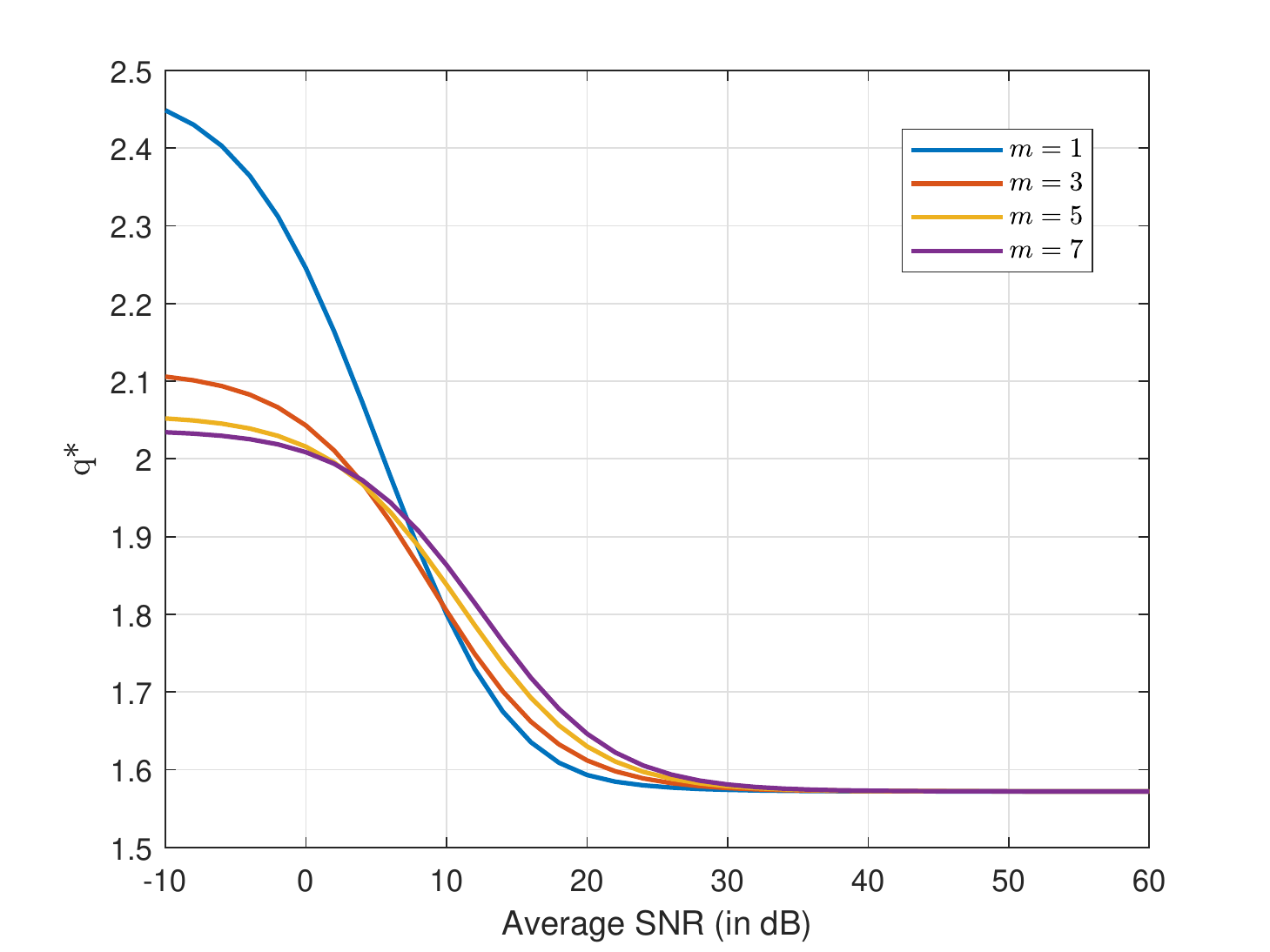}
    }
  \caption{(a) SEP curve of optimized 2-bit receiver for different $m$ and (b) corresponding $q^*_1$ }
  \label{fig:SEP_nakagami}
\end{figure}

We first investigate how SEP is affected by the quantizer structure. We consider the simple case of 2-bit 4-PAM receiver with quantization boundaries $\{-q_1,0,+q_1\}$ and constellation $\mathcal{X} = \{\pm1,\pm3\}$. Figure~\ref{fig:SEP_vs_c} depicts the SEP of the receiver as a function of $q_1$ under different spread parameter $\Omega$ (here, $\Omega SNR$ is held fixed at 10 dB). The theoretical results are generated using the analytical expression in Theorem \ref{thm1}. For completeness, Monte-Carlo simulation is also done following the ML detector in Proposition \ref{prop1}. It can be observed that a receiver with ill-placed quantization boundaries can have worse performance than a receiver equipped with an optimal quantizer even if the former has higher $\Omega$ than the latter. This emphasizes the importance of proper quantizer design in communication systems. Moreover, varying $\Omega$ does not change the optimal SEP (denoted as SEP$^*$) of the system; just the position of the quantization boundary where this SEP$^*$ occurs. Thus, we can limit our analysis of SEP$^*$ to $\Omega = 1$.

The SEP$^*$ curves under different $m$ and the optimal $q_1$ (denoted as $q_1^*$) that produced those curves are shown in Figure \ref{fig:SEP_nakagami}a and \ref{fig:SEP_nakagami}b, respectively. $q_1^*$ is obtained numerically using the gradient-based \textbf{fmincon}() MATLAB function with multiple start points. We set the objective function to be minimized as the $P_{\text{e}}$ in Theorem \ref{thm1}. The optimal quantizer in Figure \ref{fig:SEP_nakagami}b, together with the ML detector, is used to generate the simulation results. Lower SEP$^*$ is observed as the value of the shape parameter $m$ is increased (i.e. less severe fading). As $m$ increases, $q^*_1$ approaches the midpoint of $+\sqrt{\mathbb{E}[|h|^2]}a$ and $+3\sqrt{\mathbb{E}[|h|^2]}a$ in the low SNR regime. This is the optimum ML decision boundary in an AWGN channel ($m \rightarrow \infty$). Moreover, it can be seen that the shape parameter $m$ does not affect the optimal placement of $q_1$ in the SNR regime where noise is practically negligible. In fact, we will show in Section \ref{section-equidistantPAM} that the optimal quantization boundaries are affected by $\Omega$ but not by $m$ in the noiseless scenario.

\begin{figure*}[t]
  \hspace*{-.8cm}
  \subfloat[\label{fig:SEP_ADC_resolution}]{
        \includegraphics[width=.525\textwidth,draft = false]{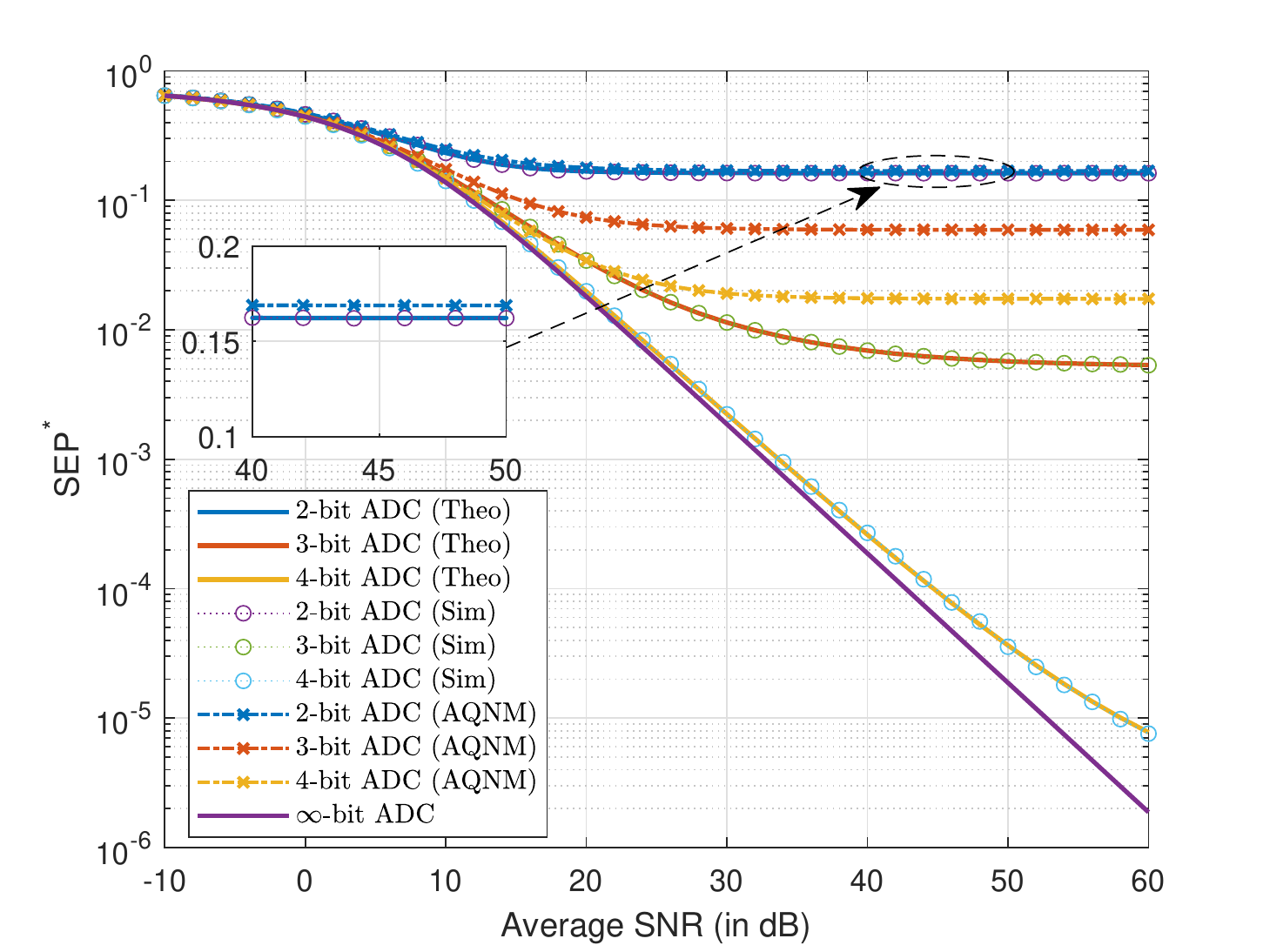}
    }
    \hspace*{-1.0cm}
    \subfloat[\label{fig:q_opt_ADC_resolution}]{
        \includegraphics[width=.525\textwidth,draft=false]{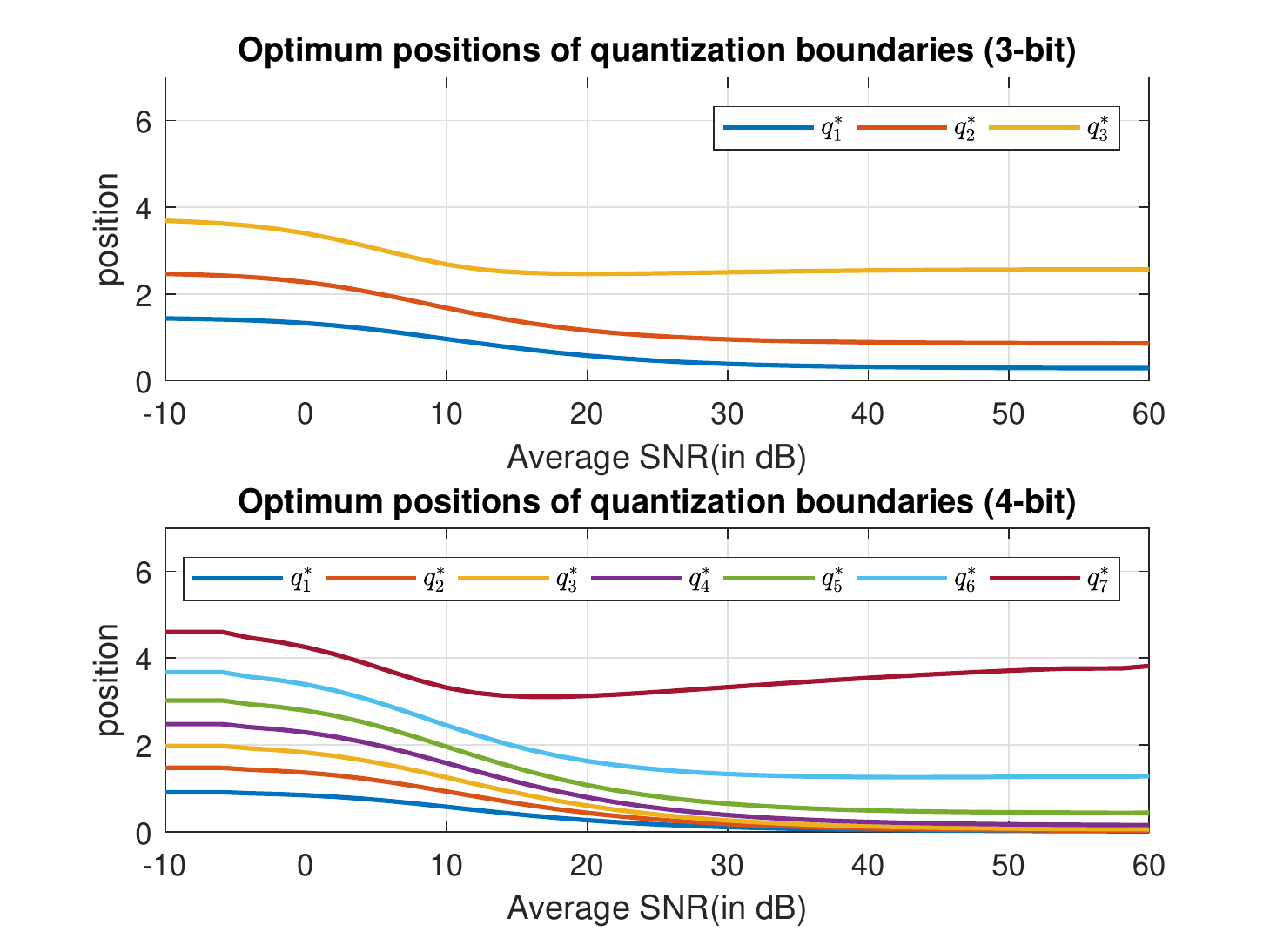}
    }
  \caption{(a) SEP$^*$ vs. SNR for different receiver resolutions $b$ (4-PAM, $m = 1$) and (b) their corresponding $\left\{q_y^*\right\}_{y=1}^{2^{b-1}-1}$ for the 3-bit and 4-bit case. The optimal quantizer for 2-bit case and $m = 1$ is already provided in Figure \ref{fig:SEP_nakagami}b.}
\end{figure*}

Next, we analyze the impact of ADC resolution on the SEP$^*$ curve when fading is Rayleigh-distributed ($m = 1$). SEP$^*$ curves for optimized 2, 3, and 4-bit 4-PAM receivers are plotted in Figure \ref{fig:SEP_ADC_resolution}. Similar to the previous numerical result, \textbf{fmincon}() is used to identify the optimal set of quantization boundaries (depicted in Figure \ref{fig:q_opt_ADC_resolution}). These quantization boundaries are used to generate the simulation results. The SEP curves when AQNM assumption is used are also superimposed in Figure \ref{fig:SEP_ADC_resolution} for comparison with our analytical model. The expression for these curves is 
\begin{equation}
    \begin{split}
    P^{\text{(AQNM)}}_{e}(SNR) =& \frac{M-1}{M}\left(1 - \sqrt{\frac{SINR}{\mathcal{E}_s + SINR}}\right),
    \end{split}
\end{equation}
where 
\begin{equation*}
     SINR = \frac{\alpha\mathcal{E}_s}{\sigma^2 + (1-\alpha)\mathcal{E}_s}.
\end{equation*}
The values for $\alpha$ are obtained from \cite[Table I]{Azizzadeh:2019}. Moreover, $\alpha \rightarrow 1$ when $b \rightarrow \infty$. A gap between our SEP expression and that of the AQNM assumption is observed in the high SNR regime. This can be attributed to the limitations of AQNM mentioned in Section \ref{section-intro}. From Figure \ref{fig:SEP_ADC_resolution}, it can be seen that there is a small loss in the uncoded error performance at $\leq 15$~dB SNR when only 3-bit quantizer is used but the SEP curve eventually reaches an error floor when SNR is increased further. Addition of a single quantization bit allows the receiver to approach the SEP of the unquantized receiver at 40 dB. Thus, during symbol detection, it is possible to use few-bit ADCs and significantly reduce the power consumption with small loss in SEP performance. However, an SEP floor is still observed at high SNR even if $b$ is increased further. Given these observations, we dedicate the next section in examining how the error floor is affected by $m$ and $b$.

\section{Analysis of Finite-Resolution $M$-PAM Receiver \\at Infinite SNR Regime}
\label{section-error-floor}

The exact average SEP of the optimum detector for $M$-PAM is analyzed in the infinite SNR regime (i.e. when $\sigma^2 = 0$). Without loss of generality, $\mathcal{E}_s = \frac{2}{M}$ so that $\sum_{i=0}^{\frac{M}{2}-1}\rho_i^2 = 1$. We first prove a lemma about the Q function.
\begin{lemma}\label{lemma2} For $\mu_A < \mu_B$, we have
\[\lim_{\sigma \rightarrow 0}Q\left(\frac{x-\mu_B}{\sigma}\right) - Q\left(\frac{x-\mu_A}{\sigma}\right) = \begin{cases}
        1, \qquad\mu_A < x < \mu_B\\
        0, \qquad otherwise
        \end{cases}. 
\]
\end{lemma}
\begin{proof}
The proof follows directly from the fact that Gaussian pdf with mean $\mu$ approaches a Dirac-Delta positioned at $\mu$ for arbitrarily small variance and the integral of a Dirac-Delta is a unit step function.
\end{proof}

Applying Lemma \ref{lemma2} on equation (\ref{eq:gen_SEP_exp}) and letting $A_{y,i}$ be the region $\left(\frac{q^2_{y-1}}{\rho^2_i},\frac{q^2_{y}}{\rho^2_i}\right)$ gives us

\begin{equation}\label{eq:SEP_noiseless}
    \begin{split}
        P_{e,\infty} \;=\;& 1 - \frac{2}{M}\sum_{y=1}^{2^{b-1}}\sum_{i=0}^{\frac{M}{2}-1}\int_{\mathcal{D}_{y,i}\cap\mathcal{A}_{y,i}}f_Z(z)\;dz,
    \end{split}
\end{equation}
where the new decision region $\mathcal{D}_{y,i} \cap \mathcal{A}_{y,i}$ is given by (\ref{eq:DA_new}). Note that the dependence of (\ref{eq:SEP_noiseless}) on $\{q_{y}\}$ is only through the integral bounds $\big[\mathcal{D}\cap\mathcal{A}\big]_{y,i}$. We introduce a non-equidistant $M$-PAM constellation, $\mathcal{X}_{\text{g}}(\rho)$, with symbols
\begin{figure*}[b]
\hrulefill
\begin{subequations}\label{eq:DA_new}
\begin{align}
    \big[\mathcal{D}\cap\mathcal{A}\big]_{y\in[1\mathrel{{.}\,{.}}\nobreak K],i\in[1\mathrel{{.}\,{.}}\nobreak \frac{M}{2}-2]}:&&      \max\left\{\frac{q_{y-1}+q_y}{\rho_i+\rho_{i+1}},\frac{q_{y-1}}{\rho_i}\right\}^2 < z& <  \min\left\{\frac{q_{y-1}+q_y}{\rho_i+\rho_{i-1}},\frac{q_{y}}{\rho_i}\right\}^2 \label{eq:DA_new_A}\\
    \big[\mathcal{D}\cap\mathcal{A}\big]_{y\in[1\mathrel{{.}\,{.}}\nobreak K],0}:&&  \max\left\{\frac{q_{y-1}+q_y}{\rho_0+\rho_1},\frac{q_{y-1}}{\rho_0} \right\}^2< z& < \frac{q^2_{y}}{\rho^2_0} \label{eq:DA_new_B}
    \\
    \big[\mathcal{D}\cap\mathcal{A}\big]_{y\in[1\mathrel{{.}\,{.}}\nobreak K+1],\frac{M}{2}-1}:&&  \frac{q^2_{y-1}}{\rho^2_{\frac{M}{2}-1}} < z& < \min\left\{\frac{q_{y-1}+q_y}{\rho_{\frac{M}{2}-1} + \rho_{\frac{M}{2}-2}},\frac{q_y}{\rho_{\frac{M}{2}-1}}\right\}^2 \label{eq:DA_new_C}.
\end{align}
\end{subequations}
\end{figure*}
\begin{equation}\label{eq:M-PAM constellation error floor}
    \begin{split}
       \mathcal{X}_{\text{g}}(\rho)
        =&\{\pm C\rho^{\frac{M}{2}-i}\}_{i = 0}^{\frac{M}{2}-1},\;\;\text{  s.t. }\;\;C^2\sum_{i=1}^{\frac{M}{2}}\rho^{2i} = 1,\;\; \rho < 1.
    \end{split}
\end{equation}
Here, $C$ is a normalizing constant and depends on $\rho$. As such, the position of the PAM symbols is controlled solely by the parameter $\rho$. We prove an optimality condition about the structure of the optimum quantizer when the transmitter uses the constellation $\mathcal{X}_{\text{g}}(\rho)$.
\begin{proposition}[\textbf{Optimality Condition of Quantizer at Infinite SNR Regime for $\mathcal{X}_{\text{g}}(\rho)$}]\label{prop2}
Given $\mathcal{X}_{\text{g}}(\rho)$ (described in equation (\ref{eq:M-PAM constellation error floor})) and $Z \sim \text{Gamma}\left(m,\frac{\Omega}{m}\right)$, the quantization boundaries $\{q_y\}_{y=2}^{2^{b-1}-1}$ of a $b$-bit symmetric quantizer should satisfy
\begin{equation}\label{eq:q_opt_condition}
    \frac{q^*_{y-1}}{q^*_{y}} = \rho
\end{equation}
to be SEP-optimal when $\sigma^2 = 0$.
\end{proposition}

\begin{proof}
See Appendix \ref{proof_prop2}.
\end{proof}
Proposition \ref{prop2} can be observed in Figure \ref{fig:q_opt_ADC_resolution} for the 3-bit and 4-bit case. The adjacent quantization boundaries for the 4-PAM constellation $\mathcal{X} = \{\pm1,\pm3\}$ have ratio $\frac{q_{y-1}}{q_y}=\frac{1}{3}$ at high SNR. The introduction of $\mathcal{X}_g(\rho)$ and Proposition \ref{prop2} will be essential later in the proofs. For now, we go back to general $M$-PAM constellation $\mathcal{X}$ and gain some intuition about the behavior of optimum SEP floor $P^{*M}_{\text{e},\infty}(m,b)$ with respect to $m$ and $b$. We define some functions, $f_{\text{L}}(m,b)$ and $f_{\text{U}}(m,b)$, such that
\[f_{\text{L}}(m,b) \leq P^{*M}_{\text{e},\infty}(m,b) \leq f_{\text{U}}(m,b).\]
For $f_{\text{U}}(m,b)$, we assume that the quantization boundaries satisfy
\begin{align}
    R = \frac{q_{y}}{q_{y-1}} = \min_{i\in[0\mathrel{{.}\,{.}}\nobreak M/2-2]}\left\{ \frac{\rho_{i+1}}{\rho_{i}}\right\}.
\end{align} 
This relationship is optimal for $M = 4$ due to Proposition \ref{prop2} but not for $M > 4$ and general $\mathcal{X}$. Using this relationship of the quantization boundaries, an error occurs whenever $|h|$ places two or more symbols inside $(0,q_1)$ or $\left(q_{2^{b-1}-1},\infty\right)$. An expression for $f_{\text{U}}(m,b)$ can be obtained as follows
\begingroup
\allowdisplaybreaks
\begin{align}\label{eq:f_U(m,b)}
    P^{*M}_{\text{e},\infty}\leq&  \frac{2}{M}\Bigg[ \sum_{i=0}^{\frac{M}{2}-2}\left(\frac{M}{2}-1-i\right)\mathbb{P}\left(|h|\rho_{i} > q_{2^{b-1}-1}\right) \nonumber\\
    &\qquad\qquad\qquad+\sum_{i=1}^{\frac{M}{2}-1}i\mathbb{P}\left(|h|\rho_i < q_1\right) \Bigg]\nonumber\\
    \leq &  \frac{2}{M}\Bigg[ \mathbb{P}\left(Z > \frac{q_{2^{b-1}-1}^2}{\rho_{\frac{M}{2}-2}^2}\right)\sum_{i=0}^{\frac{M}{2}-2}\left(\frac{M}{2}-1-i\right)\nonumber \\
    &\qquad\qquad\qquad+ \mathbb{P}\left(Z < \frac{q_1^2}{\rho_1^2}\right)\sum_{i=1}^{\frac{M}{2}-1}i  \Bigg]\nonumber\\
    =& \left(\frac{M}{4}-\frac{1}{2}\right)\left[\mathbb{P}\left(Z < \frac{q_1^2}{\rho_1^2}\right) + \mathbb{P}\left(Z > \frac{q^2_{2^{b-1}-1}}{\rho_{\frac{M}{2}-2}^2}\right)\right]
\end{align}
\endgroup
The first line follows from the fact that the error terms with exactly $n$ symbols inside a quantization region will occur $n$ times in the expression. The inequality is due to the assumption on the quantization boundaries. The second line follows from letting $Z = |h|^2$ and noting that
\begin{align*}\mathbb{P}\left(Z < \frac{q_1^2}{\rho_1^2}\right) = \max_{i \ne 0}\left\{\mathbb{P}\left(Z < \frac{q_1^2}{\rho_i^2}\right)\right\}
\end{align*}
and
\begin{align*}
\mathbb{P}\left(Z > \frac{q_{2^{b-1}-1}^2}{\rho_{\frac{M}{2}-2}^2}\right) = \max_{ i\ne\frac{M}{2}-1}\left\{\mathbb{P}\left(Z > \frac{q_{2^{b-1}-1}^2}{\rho_i^2}\right)\right\}.
\end{align*}
The third line follows from evaluating the summation terms. Equality between $P^{*M}_{\text{e},\infty}(m,b)$ and $f_{\text{U}}(m,b)$ is achieved when $M = 4$.
For $f_{\text{L}}(m,b)$, we simply use the expression in (\ref{eq:f_U(m,b)}) but replace the coefficient $\left(\frac{M}{4}-\frac{1}{2}\right)$ with $\frac{2}{M}$. That is,
\begin{equation}\label{eq:f_L(m,b)}
    f_{\text{L}}(m,b) = \frac{2}{M}\left[\mathbb{P}\left(Z < \frac{q_1^2}{\rho_1^2}\right) + \mathbb{P}\left(Z > \frac{q^2_{2^{b-1}-1}}{\rho_{\frac{M}{2}-2}^2}\right)\right].
\end{equation}
We consider four scenarios in the remainder of this section. The analysis on the behavior of the error floor with increasing $b$ or $m$ when the receiver is equipped with \textbf{(A)} SEP-optimal quantizer or a \textbf{(B)} uniform quantizer is first presented in Section \ref{section-equidistantPAM}. We then show in Section \ref{section-jointopt} that, for $M \geq 4$, error floor can be eliminated by \textbf{(C)} allowing the transmitter and receiver to jointly optimize the constellation and quantizer. There are some limitations, however, when \textbf{(D)} joint optimization of constellation and quantizer is restricted to uniform quantization.

\subsection{Optimizing Quantizer for an Equidistant $M$-PAM Constellation}\label{section-equidistantPAM}

In this subsection, we use $\mathcal{X}_{\text{eq}}(\rho) = \{\pm(2i+1)\rho\}_{i=0}^{\frac{M}{2}-1}$ for the constellation of equidistant $M$-PAM. The parameter $\rho$ controls the power of the transmitted constellation. To analyze the error floor behavior, we first define a mathematical notation for asymptotic equivalence.
\begin{definition}\label{def:asymp_eq}
We use $\sim_{x}^a$ to denote asymptotic equivalence. We say that 
\begin{align}
    f(x)\;\sim^{a}_{x}\;g(x) \iff \underset{x\rightarrow a}{\lim}\frac{f(x)}{g(x)} = 1.
\end{align} 
\end{definition}
This definition readily extends to multiple variables. For example, $f(x,y)\;\sim^{a,b}_{x,y}\;g(x,y) \iff \underset{x\rightarrow a}{\lim}\underset{y\rightarrow b}{\lim}\frac{f(x,y)}{g(x,y)} = 1$. The next two theorems establish the asymptotic behavior of the optimum error floor when equidistant $M$-PAM constellation is used.

\begin{theorem}[\textbf{$b$-bit SEP-optimal Non-uniform Quantizer, Equidistant $M$-PAM}]\label{thm2}
For any $M \geq 4$ equidistant PAM constellation and sufficiently large $b$, the SEP floor of a $b$-bit $M$-PAM receiver goes down at an exponential rate with increasing $m$ and double exponential rate with increasing $b$  if SEP-optimal quantizer is used. More formally, 
\begin{equation}\label{eq:asymp_behavior_nonunif}
P^{*M}_{\text{e},\infty}(m,b) =O\left(2^{-\left[(2^b-1)m-b\right]}\right),
\end{equation}
if
\begin{equation*}
    b > \log_2\Bigg[\frac{\log_2\left(\left[\frac{M-3}{3}\right]^2\right)}{\log_2\left(\left[\frac{M-1}{M-3}\right]\right)} +2\Bigg]+1.
\end{equation*}
\end{theorem}

\begin{proof}
See Appendix \ref{proof_thm2}.
\end{proof}

\begin{theorem}[\textbf{$b$-bit SEP-optimal Uniform Quantizer, Equidistant $M$-PAM}]\label{thm3}
For any $M \geq 4$ equidistant PAM constellation and sufficiently large $b$, the SEP floor of a $b$-bit $M$-PAM receiver goes down at an exponential rate with increasing $m$ or increasing $b$ if optimized uniform quantizer is used. More formally, 
\begin{equation}\label{eq:asymp_behavior_unif}
    P^{*M}_{\text{e},\infty}(m,b) \;=\; O\left(2^{-2bm}\right),
\end{equation}
if
\begin{equation*}
    b > \log_2\left(\frac{M-3}{3}+1\right) + 1.
\end{equation*}
\end{theorem}
\begin{proof}
See Appendix \ref{proof_thm3}.
\end{proof}

 \begin{figure*}[t]
  \subfloat[\label{fig:ErrorFloor_Nakagami_bits}]{
        \includegraphics[width=.48\textwidth,draft = false]{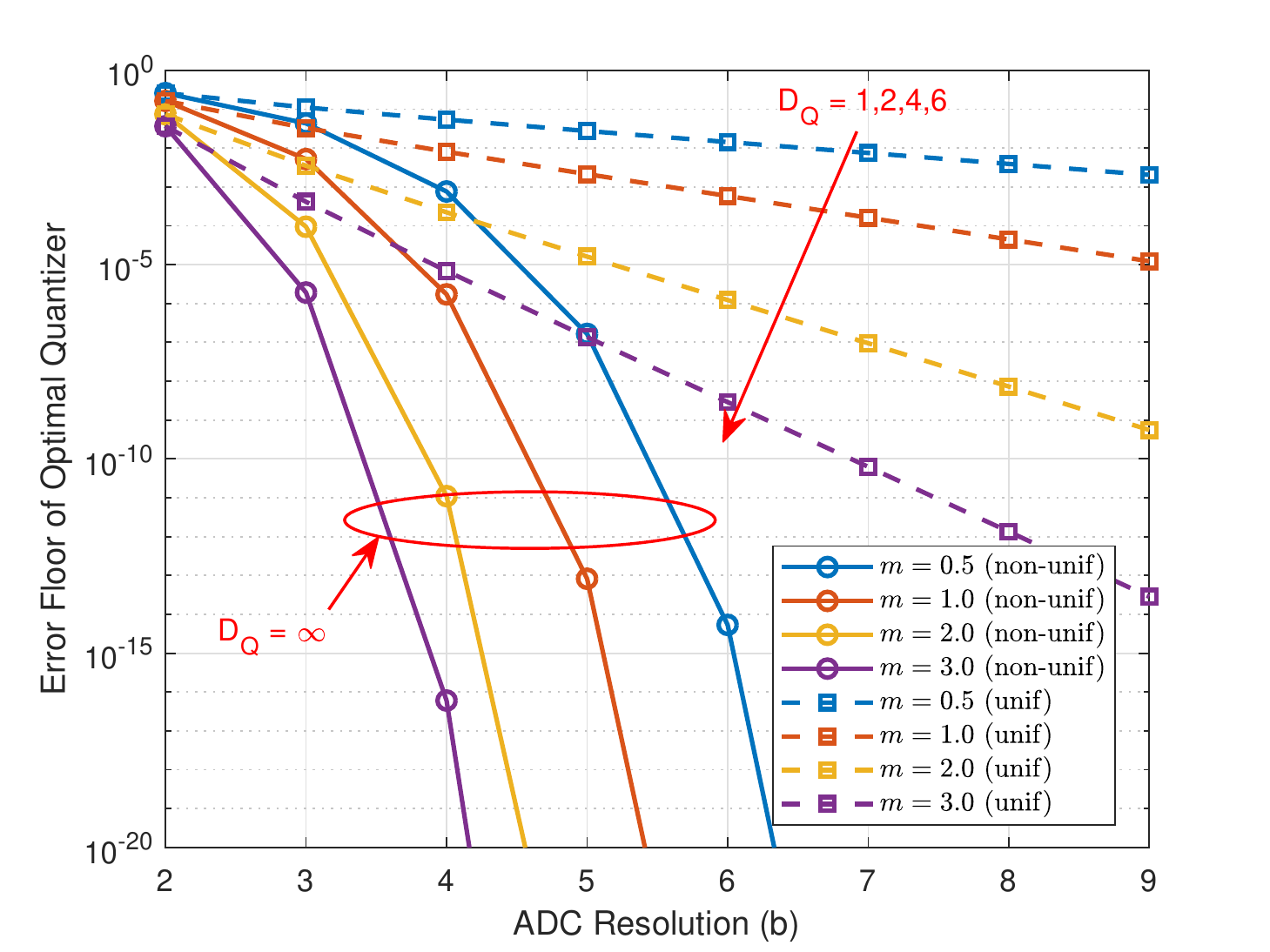}
    }
    \subfloat[\label{fig:ErrorFloor_Nakagami_Shape}]{
\includegraphics[width=.48\textwidth,draft=false]{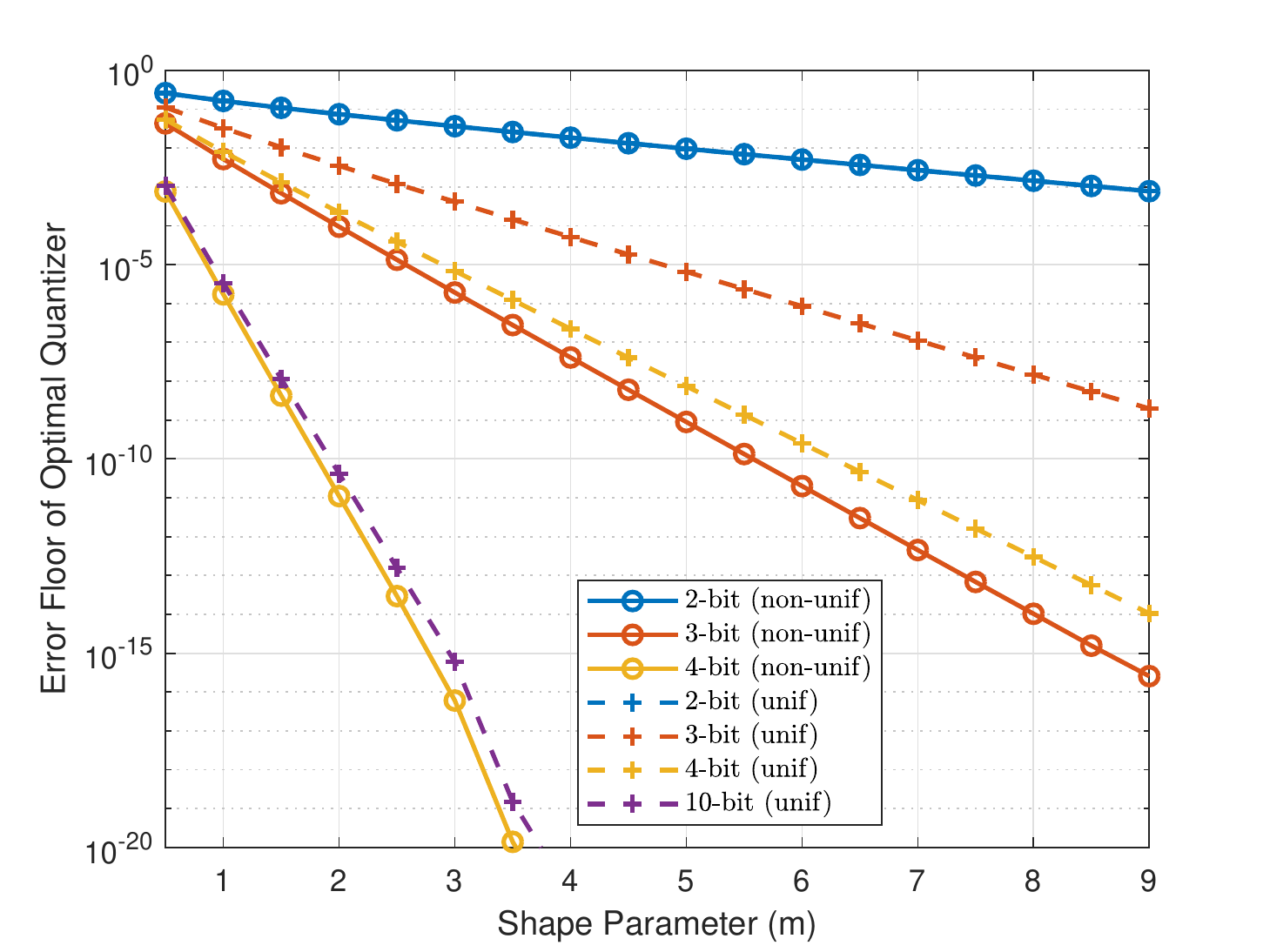}
    }
  \caption{SEP Floor of Optimal Quantizer vs.  (a) ADC resolution $b$ and (b) Shape Parameter $m$ (4-PAM)}
\end{figure*}

\textit{\noindent \textbf{Remark: } The design of the optimal quantizer for the noiseless fading environment is dependent of the spread $\Omega$ and ADC resolution $b$ but is independent of the shape $m$ (See (\ref{eq:optimal_q1_noiseless}) and (\ref{eq:delta_opt_standard_unif})). This is consistent with what was observed in Figure $\ref{fig:SEP_nakagami}$b. However, the lowest achievable SEP becomes dependent of $m$ and $b$ but is independent of $\Omega$.}

One notable difference between the two theorems is how the error floor behaves with the number of quantization bits. To illustrate this, the lowest achievable SEP floors of the optimum 4-PAM receiver for different $m$ are plotted as a function of the ADC resolution $b$ in Figure \ref{fig:ErrorFloor_Nakagami_bits}. Equations (\ref{eq:f_U(m,b) nonunif}) and (\ref{eq:f_U(m,b) unif}) are used to generate the results for non-uniform and uniform quantizer, respectively. Note that $f_\mathrm{U}(m,b)$ coincides with $f_\mathrm{L}(m,b)$ for $M = 4$ so these plots are exact. A performance measure for quantized systems with error floor was introduced in \cite{Mezghani:2008} which relates the outage performance of the receiver to the number of its quantization bits. This measure is given by
\begin{equation}\label{eq:diversity order quantized}
    \text{D}_Q = \underset{b \rightarrow \infty}{\lim} -\frac{\log_2 P_{e,\infty}(b)}{b},
\end{equation}
where $P_{e,\infty}(b)$ is the error probability as a function of $b$ in the noiseless case. Analyzing the asymptotic behavior of SEP-optimal non-uniform and uniform quantization gives us $\text{D}_Q$ values of $+\infty$ and $2m$, respectively. While increasing $b$ lowered the error floors for both SEP-optimal uniform and non-uniform quantization, larger improvements are observed in the latter due to the double exponential dependence of the SEP floor on quantization bits. This improvement, however, is not captured by the $\text{D}_Q$ metric proposed in \cite{Mezghani:2008} which hints about the limited applicability of the said metric. 
Figure \ref{fig:ErrorFloor_Nakagami_Shape} depicts the lowest achievable SEP floor of optimum 4-PAM receiver for different ADC resolutions as a function of shape parameter $m$. $\log(P_{e,\infty})$ has a downward sloping trend to an increase in $m$. The slope becomes steeper when higher ADC resolution is employed and/or when there is a design shift from uniform to non-uniform quantizer. In fact, in terms of error floor reduction in the 4-PAM case, a 4-bit SEP-optimal non-uniform quantizer even outperforms a 10-bit SEP-optimal uniform quantizer. 
\subsection{Joint Optimization of Quantizer and $M$-PAM Constellation}\label{section-jointopt}
We saw in the previous subsection that error floor cannot be eliminated with finite ADC resolution and fixed constellation. Can we do better if statistical CSI is granted at the transmitter? We assume that the transmitter can adjust its constellation $\mathcal{X}$ based on the statistics of the channel under the constraint $\sum_{i=0}^{\frac{M}{2}-1}\rho_i^2 = 1$. We prove the following lemmas about the SEP of $b$-bit $M$-PAM receiver under this scenario. Lemma \ref{lemma3} states that the error floor of finite-resolution $M$-PAM equipped with SEP-optimal quantizer can be removed if $2^b > M-2$. On the other hand, Lemma \ref{lemma4} says that if we confine the quantizer to have a uniform structure, we can remove the error floor for $M = 4$ and $b > 1$.

\begin{lemma}\label{lemma3}
Suppose $P^{M}_{e,\infty}\left(m,b,q(\rho),\mathcal{X}\left(\rho\right)\right)$ is the SEP of $b$-bit $M$-PAM receiver when the channel is subjected to Nakagami-$m$ fading at infinite SNR regime ($\sigma^2 = 0$). The quantization boundaries, $q(\rho)$, and $M$-PAM constellation, $\mathcal{X}(\rho)$, are dependent on some parameter $\rho > 0$. Then, for any $\epsilon > 0$ and $2^b > M - 2$, there exists $\delta > 0$, $\mathcal{X}(\rho)$, $q(\rho)$ such that $\rho < \delta$ and $P^{M}_{e,\infty}\left(\cdot\right) < \epsilon$.
\end{lemma}

\begin{proof}
Consider the constellation $\mathcal{X}_{\text{g}}(\rho)$ given in (\ref{eq:M-PAM constellation error floor}). The optimal quantization boundaries should satisfy Proposition \ref{prop2}. Thus, $q_y\rho^{y-1} = q_1$ and we only need to optimize on $q_1$. Then, $P^{M}_{e,\infty}$ can be upper bounded by equation (\ref{eq:f_U(m,b)}).
\begin{align}\label{eq:errorfloor M-PAM shaping}
    P^{M}_{e,\infty} \leq& \left(\frac{M}{4}-\frac{1}{2}\right)\left[\mathbb{P}\left(Z < \frac{q_1^2}{C^2\rho^{M-2}}\right) +  \mathbb{P}\left(Z > \frac{q^2_{1}}{C^2\rho^{2^b}}\right)\right]\nonumber\\
    =&
    \left(\frac{M}{4}-\frac{1}{2}\right)\frac{\gamma\left(m,\frac{m}{\Omega}\frac{q_1^2}{C^2\rho^{M-2}}\right)+\Gamma\left(m,\frac{m}{\Omega}\frac{q^2_{1}}{C^2\rho^{2^b}}\right)}{\Gamma(m)},
\end{align}
where $\gamma(m,x) = \int_{0}^xt^{m-1}e^{-t^2}\;dt$ is the lower incomplete gamma function. The upper bound of $P^{M}_{e,\infty}$ can be set to an arbitrarily small value if we can find a function $q_1(\rho) > 0$ such that 
\begin{equation*}
    \begin{split}
         \lim_{\rho\rightarrow 0}\;\;\frac{q_1^2\left(\rho\right)}{C^2\rho^{M-2}} = 0\quad \text{and}\quad \lim_{\rho\rightarrow 0}\;\;\frac{q_1^2\left(\rho\right)}{C^2\rho^{2^b}} = \infty.
    \end{split}
\end{equation*}
A solution exists if $\underset{\rho\rightarrow 0}{\lim}\;\rho^{M-2-2^{b}} = \infty$ which is satisfied when $2^b > M-2$. One such family of functions is $q_1\left(\rho\right) = \sqrt{C^2\rho^a}$ where $a \in (M-2,2^b)$.
\end{proof}

\begin{lemma}\label{lemma4}
Suppose $P^{M,\text{unif}}_{e,\infty}\left(m,b,\Delta_q(\rho),\mathcal{X}\left(\rho\right)\right)$ is the SEP of $M$-PAM receiver equipped with $b$-bit uniform quantizer when the channel is subjected to Nakagami-$m$ fading at infinite SNR regime ($\sigma^2 = 0$). The quantization step, $\Delta_q(\rho)$, and PAM constellation, $\mathcal{X}(\rho)$, are dependent on some parameter $\rho > 0$. Then, for any $\epsilon > 0$, $M = 4$, and $b > 1$, there exists $\delta > 0$, $\mathcal{X}(\rho)$, $\Delta_q(\rho)$ such that $\rho < \delta$ and $P^{4,unif}_{e,\infty}\left(\cdot\right) < \epsilon$.
\end{lemma}

\begin{proof}
Consider the constellation $\mathcal{X}_{\text{g}}(\rho)$ given in (\ref{eq:M-PAM constellation error floor}). Following the approach in Lemma \ref{lemma3}, an upper bound of the error probability of SEP-optimal uniform quantization is
    \begin{align}\label{eq:errorfloor M-PAM shaping_unif}
        P^{M,\text{unif}}_{e,\infty} \leq&
          \left(\frac{M}{4}-\frac{1}{2}\right)\Bigg[ \frac{\gamma\left(m,m\frac{\Delta_q^2}{C^2\rho^{M-2}}\right)}{\Gamma(m)}\nonumber\\
          &\qquad\qquad\qquad+\frac{\Gamma\left(m,m\frac{(2^{b-1}-1)^2\Delta^2_{q}}{C^2\rho^{4}}\right)}{\Gamma(m)}\Bigg].
    \end{align}
    The upper bound of $P^{M,unif}_{e,\infty}$ can be set to an arbitrarily small value if we can find a function $\Delta_q(\rho) > 0$ such that 
    \begin{equation*}
        \begin{split}
             \lim_{\rho\rightarrow 0}\;\;\frac{\Delta_q^2\left(\rho\right)}{C^2\rho^{M-2}} = 0\quad \text{and}\quad \lim_{\rho\rightarrow 0}\;\;\frac{(2^{b-1}-1)^2\Delta^2_{q}\left(\rho\right)}{C^2\rho^{4}} = \infty,
        \end{split}
    \end{equation*}
and a solution exists if $\underset{\rho\rightarrow0}{\lim}\;\rho^{M-6} = \infty$ which is satisfied when $M < 6$. Since we are considering $M \geq 4$ is a power of 2, this is satisfied by 4-PAM and $2^{b-1}-1 > 0$ (or $ b > 1$). 
\end{proof}
Although we proved that the SEP can be made arbitrarily small for $M = 4$ in the uniform quantizer case, we conjecture that error floor will be present when $M > 4$. Nonetheless, we analyze in the next section the decay exponent of SEP-optimal uniform quantization if $M = 4$.

Throughout this section, we have only discussed optimum quantization and error floors at infinite SNR regime. The following corollary about $\rho$ extends Lemma \ref{lemma3} and Lemma \ref{lemma4} to high SNR regime with finite SNR. To be more precise, we can have a vanishing error probability at arbitrarily small $\sigma$ if the parameter $\rho$ in Lemma \ref{lemma3} and \ref{lemma4} is also made arbitrarily small. 

\begin{corollary}\label{corollary 1}
Suppose $P^{M}_{e}\left(m,b,q(\rho),\mathcal{X}\left(\rho\right),\sigma\right)$ and $P^{M,unif}_{e}\left(m,b,q(\rho),\mathcal{X}\left(\rho\right),\sigma\right)$ are the SEP of $b$-bit $M$-PAM receiver at high SNR regime (arbitrarily small $\sigma$) for non-uniform and uniform quantization, respectively. Then, for any $\epsilon > 0$ and $2^b > M - 2$, there exists $\delta > 0$, $\mathcal{X}(\rho)$, $q(\rho)$ such that $\sqrt{\rho^2+\sigma^2} < \delta$ and $P^{M}_{e,\infty}\left(\cdot\right) < \epsilon$. Similarly, for any $\epsilon > 0$ and $b > 1$, there exists $\delta > 0$, $\mathcal{X}(\rho)$, $\Delta_q(\rho)$ such that $\sqrt{\rho^2+\sigma^2} < \delta$ and $P^{4,unif}_{e,\infty}\left(\cdot\right) < \epsilon$.
\end{corollary}

\begin{proof}
Since our SEP expression for finite $SNR$ given in Theorem \ref{thm1} is a continuous function, then assuming that the corollary does not hold will contradict Lemma \ref{lemma3} and Lemma \ref{lemma4}. 
\end{proof}

\section{Impact of Quantization on Diversity Order of\\ $M$-PAM:The case of statistical CSIT}
\label{section-diversity-order}

We have shown the existence of an irreducible SEP floor with ML detection of equidistant $M$-PAM transmission with CSIR and having knowledge of statistical CSI at the transmitter can eliminate this SEP floor by allowing it to optimize the constellation. In this section, we analyze the decay exponent of finite-resolution $M$-PAM receiver with jointly-optimized constellation and quantizer for a Nakagami-$m$ fading channel. Without loss of generality, we restrict our analysis to $\Omega = 1$. The decay exponent is defined as 
\begin{equation}\label{eq:DVO_definition}
    \begin{split}
    \text{DVO} =& -\underset{SNR\rightarrow \infty}{\lim} \frac{\log P_e\left(SNR\right)}{\log SNR} = \underset{\sigma^2\rightarrow 0}{\lim} \frac{\log P_e\left(\frac{1}{\sigma^2}\right)}{\log \sigma^2}.
    \end{split}
\end{equation}
We use the term DVO to denote diversity order, which is the asymptotic slope of the error probability as a function of SNR. Decay exponent and diversity order is used interchangeably in this paper. In addition, we use the special symbol $\doteq$ to denote \emph{exponential equality}, a concept introduced in Diversity-Multiplexing Trade-off (DMT) analysis \cite{Zheng:2003}. We say that $f(SNR) \doteq SNR^b$ (or $f\left(\frac{1}{\sigma^2}\right) \doteq \left(\frac{1}{\sigma^2}\right)^b$) if
\begin{equation*}
    \underset{SNR\rightarrow\infty}{\lim} \frac{\log f(SNR)}{\log SNR} =b \quad \iff \quad  \underset{\sigma^2\rightarrow 0}{\lim} \frac{\log f\left(\frac{1}{\sigma^2}\right)}{\log \sigma^2} = -b.
\end{equation*}
The transmitter sends symbol $x \in \mathcal{X} = \{\pm\rho_i\}_{i=0}^{\frac{M}{2}-1}$ and its knowledge of statistical CSI allows it to strategically allocate energy. Before we head straight to the derivation of the decay exponent, we first present two key properties of exponential equality. Lemmas \ref{lemma5}.i and \ref{lemma5}.ii show the summation property and scaling invariance property of exponential equality, respectively.
\begin{lemma}\label{lemma5}
    Suppose we have $f(SNR) \doteq SNR^{d}$ and $f_i(SNR) \doteq SNR^{d_i}$. Then,
    \begingroup
    \allowdisplaybreaks
    \begin{align*}
    (i) & && \text{$\sum_{i=1}^Nf_i(SNR) \doteq SNR^{d_{max}}$, where $d_{max} = \max_{i\in[1,N]}{d_i}$}\\
    (ii) & && \text{For any $\alpha > 0$, $\alpha f(SNR) \doteq SNR^d$} 
    \end{align*}
    \endgroup
\end{lemma}
\begin{proof}
 See proof of \cite[Lemma 2]{Gayan:2020}
\end{proof}
Using the properties of exponential equality stated in Lemma \ref{lemma5}, we reduce the exact SEP to a simpler but exponentially equivalent expression in Lemma \ref{lemma6}. We then present a function $f_0\left(\frac{1}{\sigma^2},\rho(\sigma),A,B,C\right)$ in Lemma \ref{lemma7} which has a DVO expressed in terms of $A$,$B$, and $C$.
\begin{lemma}\label{lemma6}
    Let $P^{M}_e\left(\frac{1}{\sigma^2}\right) $ be the error probability of $M$-PAM with constellation $\{\pm\rho_i\}^{\frac{M}{2}-1}_{i=0}$ (where $|\rho_i| < |\rho_{i+1}|$) over Nakagami-$m$ fading, where $m\geq \frac{1}{2}$. Then,
    \begin{equation*}
        \begin{split}
           P^M_e\left(\frac{1}{\sigma^2}\right) \doteq \left[\frac{\sigma^2}{\rho_0^2}\right]^{m}+ \sum_{i = 0}^{\frac{M}{2}-1}\sum_{n=0,n\ne i}^{\frac{M}{2}-1}\mathbb{P}(\hat{x} = +\rho_n|x = +\rho_i).
        \end{split}
    \end{equation*}
\end{lemma}
\begin{proof}
    To prove this, we expand the SEP expression as follows
    \begingroup
    \begin{align}\label{eq:P_e_lemma6} P^M_e\left(\frac{1}{\sigma^2}\right) =& \frac{2}{M}\sum_{i = 0}^{\frac{M}{2}-1}\mathbb{P}(\hat{x} \ne +\rho_i|x = +\rho_i)\nonumber \\
            =& \frac{2}{M}\sum_{i = 0}^{\frac{M}{2}-1}\Bigg[\mathbb{P}(y \leq 0|x = +\rho_i) \nonumber\\
            &\qquad\qquad+ \sum_{n=0,n\ne i}^{\frac{M}{2}-1}\mathbb{P}(\hat{x} = +\rho_n|x = +\rho_i)\Bigg]\nonumber\\
            \doteq& \sum_{i = 0}^{\frac{M}{2}-1}\left[\frac{\sigma^2}{\rho_i^2}\right]^m + \sum_{i = 0}^{\frac{M}{2}-1}\sum_{n=0,n\ne i}^{\frac{M}{2}-1}\mathbb{P}(\hat{x} = +\rho_n|x = +\rho_i),
\end{align}
\endgroup
where we used the asymptotic SEP expression of a BPSK system $\{-\rho_i,+\rho_i\}$ over Nakagami-$m$ fading at high SNR \cite{Wang:2003} for $\mathbb{P}(y \leq 0| x = \rho_i)$. This substitution is valid since an error is committed in a BPSK system if the received signal falls in the negative region given the positive symbol is transmitted. This corresponds to an ADC output $y \leq 0$. The coefficients of each term in the right-hand side are dropped due to Property (ii) of Lemma \ref{lemma5}. To simplify the first summation term, we observe the relative growth of $\rho_0$ compared to symbol $\rho_i,\;i > 0$. 
\begin{equation*}
    \begin{split}
   \underset{\sigma^2\rightarrow 0}{\lim} \frac{\left[\frac{\rho_i^2}{m\sigma^2}\right]^{-m}}{\left[\frac{\rho_0^2}{m\sigma^2}\right]^{-m}}
    =& \left[\frac{\rho_0^2}{\rho_i^2}\right]^{m}.
    \end{split}
\end{equation*}
Since $0 < \rho_0 < \rho_i$, then $\left[\frac{\rho_0^2}{\rho_i^2}\right]^m$ is finite. Using Lemma \ref{lemma5}.i on (\ref{eq:P_e_lemma6}) completes the proof.
\end{proof}

\begin{figure*}[t]
  \hspace*{-.5cm}
  \subfloat[\label{fig:SEP_divOrder}]{
        \includegraphics[width=.52\textwidth,draft = false]{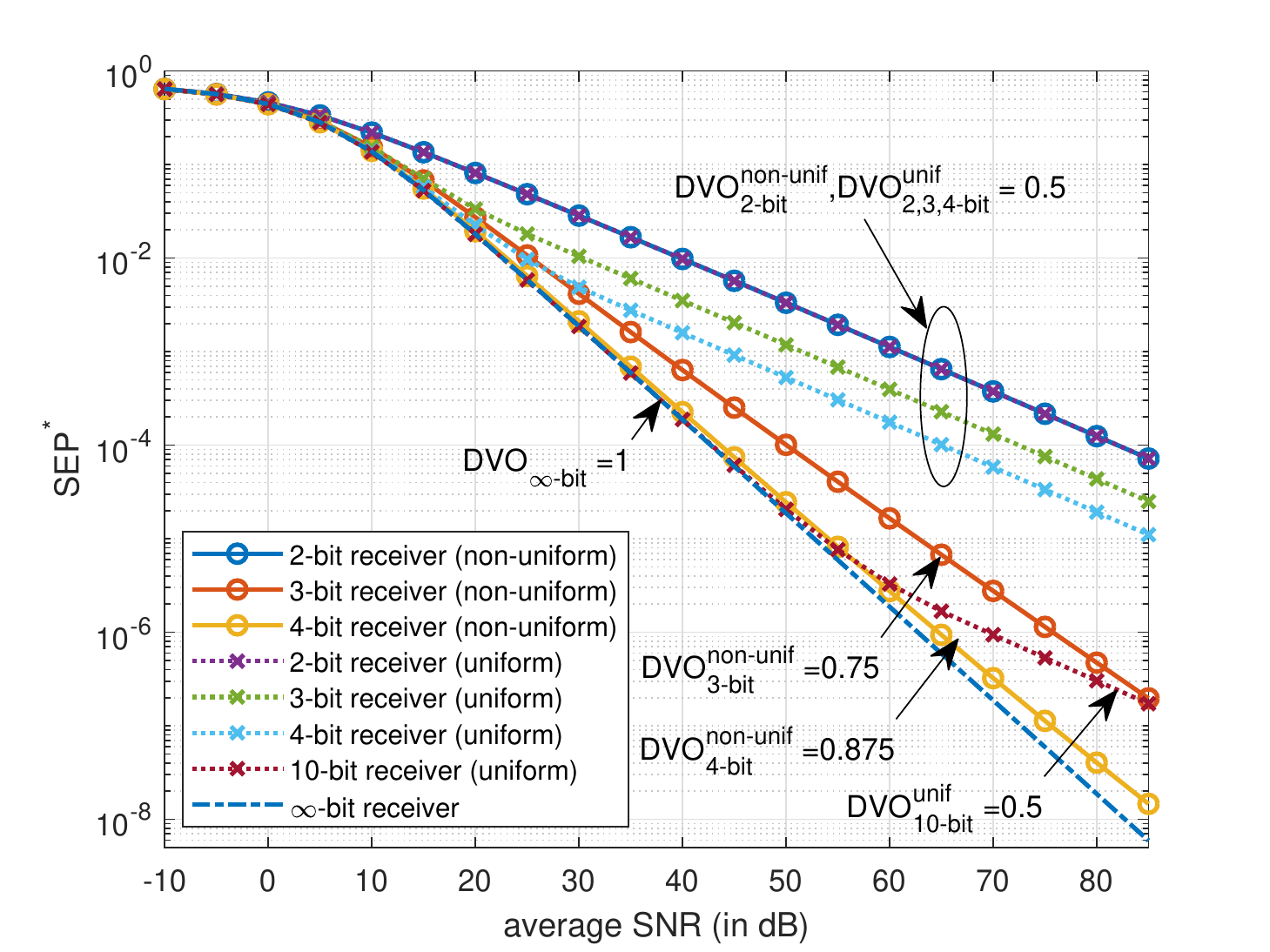}
    }
    \hspace*{-.75cm}
    \subfloat[\label{fig:q_opt_DivOrder}]{
        \includegraphics[width=.52\textwidth,draft=false]{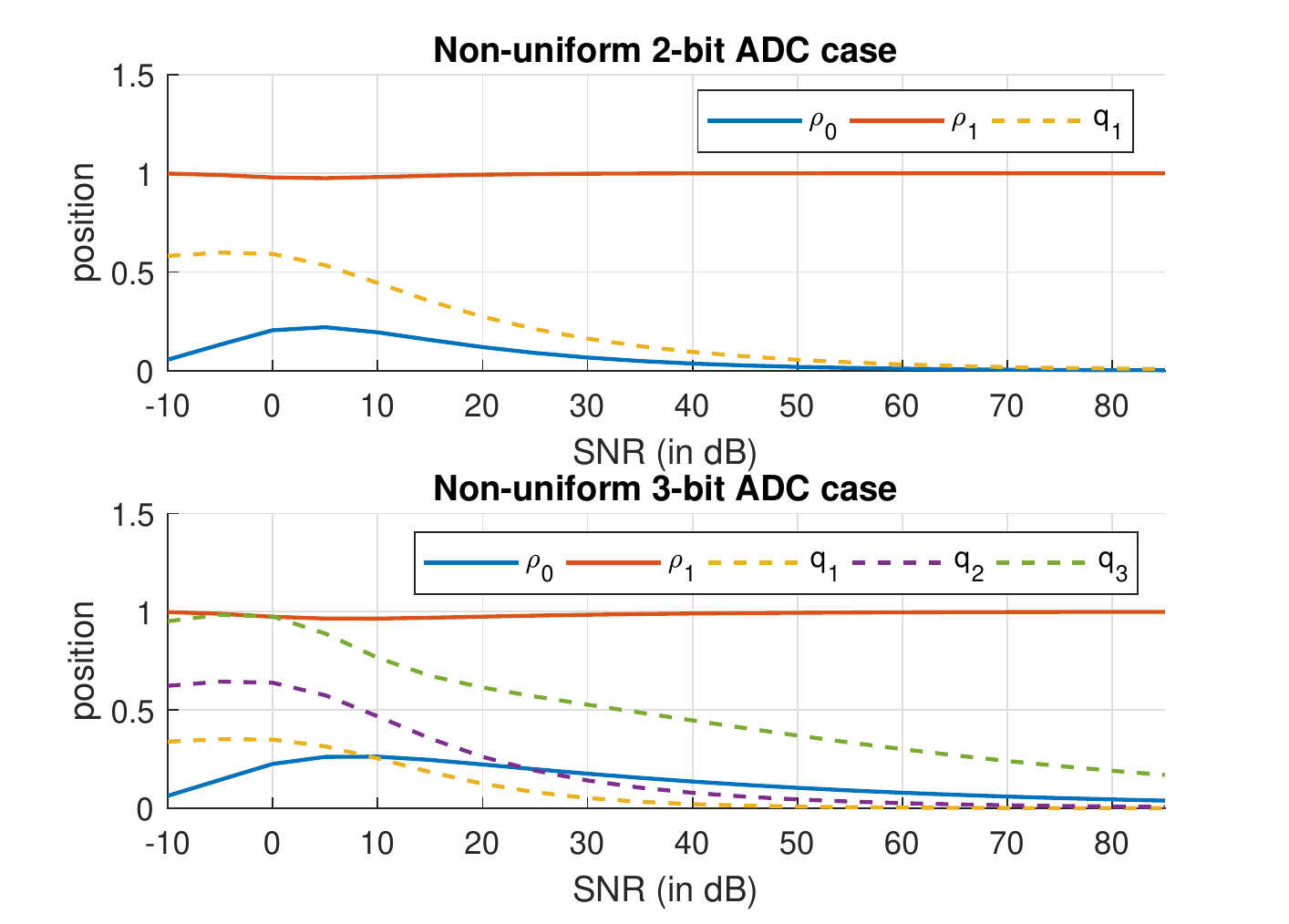}
    }
  \caption{(a) Optimum SEP of low-resolution 4-PAM receiver with joint quantizer and constellation optimization. Theoretical DVO values are also shown in the plot. (b) Numerically-computed quantizer and constellation for non-uniform 2-bit, 3-bit ADCs.}
\end{figure*}

\begin{lemma}\label{lemma7}
    Suppose we have a function $f_0\left(\frac{1}{\sigma^2},\rho(\sigma),A,B,C\right)$ defined as
    \begin{equation}\label{eq:f_0}
    f_0\left(\frac{1}{\sigma^2},\rho(\sigma),A,B,C\right) =\left[\frac{\sigma^2}{\{\rho\left(\sigma\right)\}^B}\right]^{C} +\{\rho\left(\sigma\right)\}^{A}
    \end{equation}
for some $A,B,C,\sigma > 0$, and $\rho(\sigma) > 0$ is a function that depends on $\sigma$. Then, 
\begingroup
\allowdisplaybreaks
\begin{align*}
    (i)& &&\underset{\sigma \rightarrow 0}{\lim}\;\;\frac{\rho^{*2}\left(\sigma\right)}{\sigma^2} = \infty\\
    (ii) & &&f_0\left(\frac{1}{\sigma^2},\rho^*(\sigma),A,B,C\right) \doteq \left[\frac{1}{\sigma^2}\right]^{-\frac{AC}{A+BC}}
\end{align*}
\endgroup
if $C < A+BC$. $\rho^*(\sigma)$ is the optimum choice of $\rho$ for some value of $\sigma$. 
\end{lemma}

\begin{proof}
For simplicity of notation, we refer to $f_0\left(\frac{1}{\sigma^2},\rho(\sigma),A,B,C\right)$ as $f_0(\cdot)$. We can get an expression that relates $\sigma^2$ and the optimum $\rho$ by differentiating $f_0(\cdot)$ with respect to $\rho$ and equating the result to 0. That is,
\begin{equation*}
    \begin{split}
        \frac{\partial f_0(\cdot)}{\partial \rho} = 0 =&-\frac{BC}{\rho}\left[\frac{\sigma^2}{\rho^B}\right]^{C} + A\rho^{A-1},
    \end{split}
\end{equation*}
which implies that
\begin{equation}
\label{eq:rho_0_opt}
    \begin{split}
        \rho^*\left(\sigma\right) = \left[\frac{BC}{A}\right]^{\frac{1}{A+BC}}\left[\sigma^2\right]^{\frac{C}{A+BC}}
    \end{split}
\end{equation}
It can be observed that the function $\rho^*(\sigma)$ decreases as $\sigma$ decreases. Moreover,
\begin{equation*}
    \begin{split}
       \underset{\sigma \rightarrow 0}{\lim}\;\frac{\rho^{*2}\left(\sigma\right)}{\sigma^2} &= \underset{\sigma \rightarrow 0}{\lim}\;\left[\frac{BC}{A}\right]^{\frac{1}{A+BC}}\left[\sigma^2\right]^{\frac{C}{A+BC}-1}=\infty
    \end{split}
\end{equation*}
since $\frac{C}{A+BC} < 1$. This proves claim (i). Substituting the expression of $\rho^*(\sigma)$ in $f_0(\cdot)$ gives
\begin{align}
   f_0\left(\frac{1}{\sigma^2},\rho^*(\sigma),A,B,C\right)  =& \left[\frac{A}{BC}\right]^{\frac{C}{A+BC}}\left[\frac{1}{\sigma^2}\right]^{-\frac{AC}{A+BC}}\nonumber\\
   &\qquad+ \left[\frac{BC}{A}\right]^{\frac{A}{A+BC}}\left[\frac{1}{\sigma^2}\right]^{-\frac{AC}{A+BC}}\nonumber\\
   \doteq& \left[\frac{1}{\sigma^2}\right]^{-\frac{AC}{A+BC}},
\end{align}
where the last line is obtained using Lemma \ref{lemma5}.i and \ref{lemma5}.ii. This completes the proof of claim (ii).
\end{proof}
Theorem \ref{thm4} and Theorem \ref{thm5} describe the diversity order of a $b$-bit $M$-PAM receiver. The former is for SEP-optimal quantizer and the latter is when we restrict the quantizer structure to be uniform. In this context, an optimized constellation and quantizer minimizes SEP. While we are only able to show that uniform quantization can only eliminate error floor for $M = 4$, the DVO for the 4-PAM case is still derived for comparison with the SEP-optimal quantizer.
\begin{theorem}[\textbf{DVO of Jointly-optimized Constellation and Non-uniform Quantizer}]\label{thm4}
Suppose the transmitter can optimize the constellation using statistical CSI. Then, the decay exponent of a $b$-bit $M$-PAM receiver equipped with SEP-optimal quantizer is
\begin{equation}\label{eq:DVO_nakagami}
    \text{DVO} = m\frac{2^{b}-M+2}{2^{b}}
\end{equation}
\noindent for Nakagami-$m$ fading and $2^b > M-2$.
\end{theorem}

\begin{proof}
See Appendix \ref{proof_thm4}.   
\end{proof}

\begin{theorem}[\textbf{DVO of Jointly-optimized Constellation and Uniform Quantizer}]\label{thm5}
Suppose the transmitter can optimize the constellation using statistical CSI. Then, the decay exponent of a $b$-bit 4-PAM receiver equipped with optimized uniform quantizer is
\begin{equation}\label{eq:DVO_nakagami_unif}
    \text{DVO} = \frac{m}{2}
\end{equation}
\noindent for Nakagami-$m$ fading and $b \geq 2$.
\end{theorem}
\begin{proof}
 See Appendix \ref{proof_thm5}.
\end{proof}

Figure \ref{fig:SEP_divOrder} depicts the SEP$^*$ curves of the 4-PAM receiver with optimized constellation and quantizer for different ADC resolutions and $m = 1$. The SEP$^*$ curves are obtained by numerically optimizing $\{\rho\}_{i=0}^{M/2-1}$ and $\{q_y\}_{y=1}^{y=2^{b-1}-1}$ in the SEP expression given in Theorem \ref{thm1} using MATLAB's \textbf{fmincon}(). The numerically-computed settings of the quantizer and constellation are shown in Figure \ref{fig:q_opt_DivOrder}\footnote{We omit the plots of the quantizer and constellation for the other cases due to page limitation.}. For the uniform quantizer case, we simply optimize over a single parameter $\Delta_q$ and replace $q_y$ with $y\Delta_q$. While no error floor is observed in this case, the SEP$^*$ decays slower for low-resolution receivers as we increase SNR. These decay exponents are given by (\ref{eq:DVO_nakagami}) and (\ref{eq:DVO_nakagami_unif}) in Theorem \ref{thm4} and Theorem \ref{thm5}, respectively. Equation (\ref{eq:DVO_nakagami}) shows a trade-off between DVO and $M$ for fixed number of quantization bits. In addition, we see that an SEP-optimal uniform quantizer has a fixed $\text{DVO}$ for all ADC resolutions in Figure \ref{fig:SEP_divOrder} unlike the non-uniform quantizer case which gradually approaches the $\text{DVO}$ of unquantized receiver as $b$ is increased. Even the 10-bit SEP-optimal uniform quantizer eventually gets a $\text{DVO}$ of $\frac{1}{2}$ at high SNR values. However, despite sharing the same $\text{DVO}$, the performance gain in using more quantization bits manifests as power offset at high SNR. We also see that $\rho_0^* \rightarrow 0$ as SNR$\rightarrow \infty$ in Figure \ref{fig:q_opt_DivOrder}. This is consistent with Corollary \ref{corollary 1}.

The above result gives further merit to the use of non-uniform quantization in low-resolution receiver designs. This result is also particularly useful for low-resolution $M$-PAM receivers since they start to approach the error floor within practical range of SNR values. Using statistical CSI to optimize the constellation eliminates this error floor albeit at a lower decay exponent. A system designer may opt to trade off reliability to reduce the power consumption. Finally, we note that unlike phase-quantized PSK receivers which achieve full diversity with sufficient quantization bits \cite{Gayan:2020}, the DVO of finite-resolution PAM receivers is strictly less than $m$.

\section{Diversity Order of Quantized SIMO Fading \\Channels}\label{section-multiantenna}
Given the diversity order results in Theorems \ref{thm4} and \ref{thm5}, we now ask the following question: \textit{Do diversity gains from increasing the number of antennas and quantization bits cumulate?} We briefly discuss the complementary roles of the number of receive antennas and the number of quantization bits in improving the diversity order of the system. We consider an extension of the receiver in Figure \ref{fig:sys_model} to the multiple antenna case. The receiver is equipped with $N_r$ receiver chains and the $i$-th ADC takes the (phase-adjusted) signal $\text{Re}\{r'_i\} = |h_i|x+\text{Re}\{w_i\}$ as input and produces the quantized observation $y_i$. In this model, $\text{Re}\{w_i\}\sim \mathcal{N}(0,\sigma^2/2)$ is the noise at the $i$-th receiver chain and $|h_i|$ is the Nakagami-distributed amplitude fading at the $i$-th receiver chain. We further assume that the transmitter can optimize its constellation using statistical CSI and the receiver chains are equipped with identical SEP-optimal quantizers. This assumption is reasonable in an i.i.d. fading environment. With channel realizations known at the receiver, the goal of the detector is to reliably recover the transmitted symbol from the quantized observations $\boldsymbol{y} = [y_1,\cdots,y_{N_r}]$. 

\begin{figure}[t]
    \centering
    \includegraphics[scale = .625]{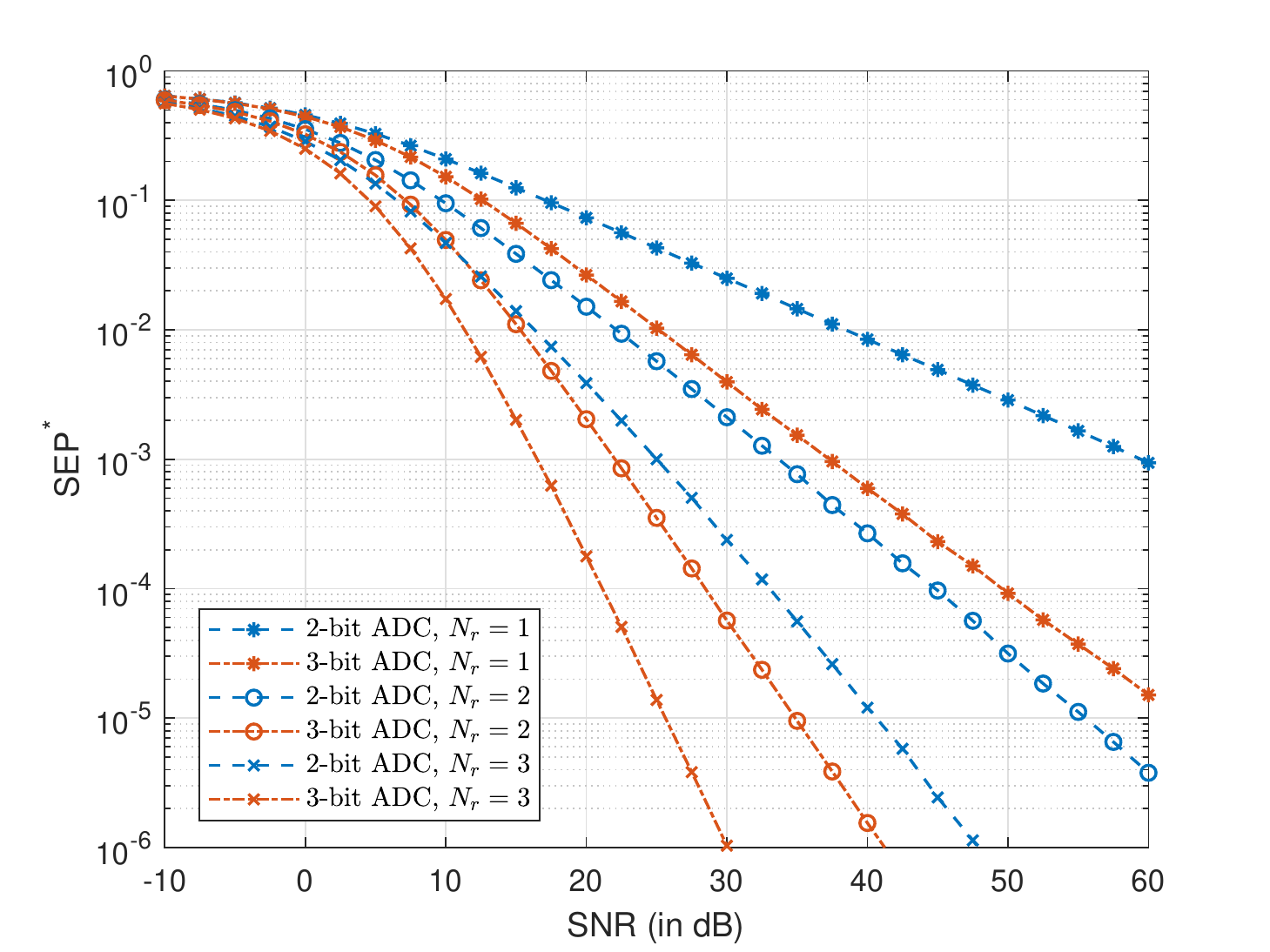}
    \caption{Simulated SEP$^*$ curve of optimized 4-PAM receiver with $N_r = \{1,2,3\}$ antennas, $b = \{2,3\}$-bits, and i.i.d. Rayleigh fading ($m = 1$). The constellation and quantizer used in the simulation are given in Figure \ref{fig:q_opt_DivOrder}.}
    \label{fig:SEP_SIMO}
\end{figure}

Suppose the transmitter uses $\mathcal{X}_g$ in (\ref{eq:M-PAM constellation error floor}) and each quantizer is designed based on Proposition \ref{prop2}. In the infinite SNR regime, an error occurs if, for all receiver chains, $|h_i|$ placed two or more symbols inside $(0,q_1)$ or $\left(q_{2^{b-1}-1},\infty\right)$. Equivalently, the SEP$^*$ at this regime can be upper bounded by
\begin{align}\label{eq:errorfloor_multiantenna}
  P_{\text{e},\infty}^{M,N_r} \leq & \Bigg\{\left(\frac{M}{4}-\frac{1}{2}\right)\Bigg[\frac{\gamma\left(m,\frac{m}{\Omega}\frac{q_1^2}{C^2\rho^{M-2}}\right)}{\Gamma(m)}\nonumber\\
  &\qquad\qquad\quad+\frac{\Gamma\left(m,\frac{m}{\Omega}\frac{q^2_{1}}{C^2\rho^{2^b}}\right)}{\Gamma(m)}\Bigg]\Bigg\}^{N_r},
\end{align}
which can be made arbitrarily small with our choice of constellation and quantizer structure. This comes from extending equation (\ref{eq:errorfloor M-PAM shaping}) in the proof of Lemma \ref{lemma3} to $N_r$ antennas. Following the approach in the proof of Theorem \ref{thm4}, the SEP of the multi-antenna $M$-PAM receiver with optimal $b$-bit quantization (denoted as $P^{M,N_r}_{\text{e}}$) can be shown to be exponentially equal to
\begingroup
        \allowdisplaybreaks
        \begin{align}\label{eq:Pe_highSNR_multiantenna}
          P^{M,N_r}_{\text{e}}\left(\frac{1}{\sigma^2}\right) \doteq& \left[\frac{\sigma^2}{\rho^{M-2}}\right]^{m N_r} +\rho^{mnN_r},
    \end{align}
\endgroup
where $n = 2^{b}-M+2$, and $2^b > M - 2$. The first term of (\ref{eq:Pe_highSNR_multiantenna}) is obtained from the asymptotic SEP of maximum ratio combining (MRC) receiver over $N_r$ i.i.d. Nakagami-$m$ fading channels \cite{Kim:2012} while the second term of (\ref{eq:Pe_highSNR_multiantenna}) comes from the fact that (\ref{eq:errorfloor_multiantenna}) is asymptotically equal to $\rho^{mnN_r}$ up to a scaling factor for arbitrarily small $\rho$. By Lemma \ref{lemma7}, the diversity order of (\ref{eq:Pe_highSNR_multiantenna}) is
\begin{align}\label{eq:DVO_SIMO}
    \text{DVO} = mN_r\frac{2^b-M+2}{2^b},\;\;\text{ if $2^b>M-2$}.
\end{align}
Thus, under the i.i.d assumption, the diversity gains from increasing the quantization bits and increasing the number of receive antennas cumulate. We demonstrate this property in Figure \ref{fig:SEP_SIMO}. The simulated results are generated using the PAM constellation and quantizer structure given in Figure \ref{fig:q_opt_DivOrder}. The receiver uses the ML rule
\begingroup
\begin{align*}
    \hat{x}^{*} &=\; \underset{x\in\mathcal{X}}{\arg \max}\;\prod_{n = 1}^{N_{r}}\left\{Q\left(\frac{q_{y_n-1} - |h_n|x}{\sqrt{\sigma^2/2}}\right) - 
         Q\left(\frac{q_{y_n} - |h_n|x}{\sqrt{\sigma^2/2}}\right)\right\}
\end{align*}
\endgroup
to recover the transmitted symbol. Huge improvement in the DVO of the SEP$^*$ curve is observed when both antenna elements and quantization bits are increased simultaneously. 

Before we conclude this section, we would like to emphasize the significance of this result. That is, we were able to show that the conventional definition of diversity order may still be a sensible metric in some fading scenarios involving low-resolution quantized receivers. This is not obvious in simplified analytical models due to the existence of irreducible quantization noise that causes an error floor. Extension of this result to the MIMO case is also worth investigating. Can we show the existence/non-existence of space-time codes and quantizer structure for MIMO systems that achieve a non-vanishing diversity order? Existence of such codes with unity code rate has been established in \cite{Ivrlac:2006} for a 2$\times$2 MIMO system with 1-bit output through exhaustive search. However, this has not been generalized to multi-bit quantization or to more than two antennas.

\section{Conclusion}
\label{section-conclusions}

\begin{table*}[t]
\renewcommand{\arraystretch}{1}
\caption{Summary of our Main Results for Different Quantizer Structures and Scenarios}
\label{tab:summary_results}
\centering
\begin{tabular}{cccc}
\hline
\textbf{ } &\textbf{Equidistant $M$-PAM Constellation} & \textbf{Optimized $M$-PAM Constellation} \\\hline\hline
\begin{tabular}{@{}c@{}}\textbf{SEP-optimal Non-uniform} \\\textbf{Quantization with $b$-bits} \end{tabular}& \begin{tabular}{@{}c@{}}$P^{*M}_{e,\infty}(m,b) = O\left(2^{-\left[(2^b-1)m-b\right]}\right)$\\ $\text{if}\;\;b > \log_2\Big[\frac{\log_2\left(\left[\frac{M-3}{3}\right]^2\right)}{\log_2\left(\left[\frac{M-1}{M-3}\right]\right)} +2\Big]+1$ \end{tabular}& DVO = $m\frac{2^b-M+2}{2^b}$, for $2^b > M - 2$
 \\\hline
\begin{tabular}{@{}c@{}}
    \textbf{SEP-optimal Uniform}\\ \textbf{Quantization with $b$-bits}
\end{tabular} & \begin{tabular}{@{}c@{}}$P^{*M}_{e,\infty}(m,b) = O\left(2^{-2bm}\right)$\\$\text{ if } b > \log_2\left(\frac{M-3}{3}+1\right) + 1$ \end{tabular} & DVO $= \frac{m}{2}$, for $M = 4$ \\\hline
\end{tabular}
\end{table*}
In this work, we analyzed the error rate of $M$-PAM receivers subjected to fading and $b$-bit quantization. Optimum decision rule and exact SEP expression for $b$-bit receiver subjected to Nakagami-$m$ fading were derived. By optimizing the quantization boundaries under the minimum SEP criterion, we showed that the SEP of unquantized $M$-PAM case be approached by a finite-resolution receiver for up to some SNR value but eventually reaches an error floor. This error floor goes down at a double exponential rate as we increase the resolution of the SEP-optimal quantizer and at an exponential rate as we increase $m$. We proved that this trend holds for $M \geq 4$. The SEP floor analysis is particularly interesting from a theoretical perspective since it presents fundamental limits of a quantizer in doing symbol detection. We also proved that the error floor can be eliminated by allowing the transmitter to shape the constellation depending on the statistical CSI. Characterization of the decay exponent of jointly-optimized $M$-PAM constellation and quantization revealed that a finite-resolution receiver has lower decay exponent than its unquantized counterpart. The decay exponent can be improved by increasing the number of quantization bits of the SEP-optimal non-uniform quantizer. We also demonstrated the complementary roles of quantization bits and number of antenna elements in improving the decay exponent. For the case of SEP-optimal uniform quantization, we only proved that the error floor can be eliminated for $M = 4$ and the decay exponent is fixed for any finite $b \geq 2$. These results are summarized in Table \ref{tab:summary_results}.


It is worth mentioning that while only SISO and SIMO $M$-PAM is considered in this work, the study provides an insightful analytical approach on how to investigate the exact error performance of a communication system equipped with SEP-optimal quantizer over a fading channel. This work also gives a link between the actual structure of the multi-bit amplitude quantizer and the average symbol error rate of the communication link over a fading environment which, to the best of our knowledge, is still missing in the literature. This connection is not apparent in simplified analytical models for quantized receivers such as AQNM. We only considered coherent $M$-PAM in this study to get some insights on the structure of SEP-optimal quantizer in one-dimensional case but design of optimal detectors and quantizers for complex-valued modulation schemes is currently being investigated. For instance, we can combine our result with that of \cite{Gayan:2020} to investigate the SEP$^*$ performance of amplitude-phase shift keying (APSK)\footnote{In some text, this modulation scheme is also known as star-QAM} receiver equipped with low-resolution polar quantizers. Our analysis of the multiple antenna case simply considered independent fading. The impact of antenna correlation on the optimal quantization and detection strategy is worth studying. Lastly, It is also of interest to analyze the capacity limits of such channel and design quantizer and signaling schemes for coded systems. 

\begin{appendices}
\section{Proof of Proposition \ref{prop1}}
\label{proof_prop1}

Suppose we received some value $y$ and we are to choose between two symbols $x_{A}$ and $x_{B}$. Then, the ML detector will choose $x_{A}$ if
\begin{align}
  \label{eq:ML_detector_rule1A}
    &Q\left(\frac{q_{y-1} - |h|x_{A}}{\sqrt{\sigma^2/2}}\right) - Q\left(\frac{q_{y} - |h|x_{A}}{\sqrt{\sigma^2/2}}\right)\nonumber \\
    &\qquad\qquad>  Q\left(\frac{q_{y-1} - |h|x_{B}}{\sqrt{\sigma^2/2}}\right) - Q\left(\frac{q_{y} - |h|x_{B}}{\sqrt{\sigma^2/2}}\right).
\end{align}
Let $\tilde{x} = x_{A}-x_{B}$, $\Delta_{y} = q_{y} - q_{y-1}$, $z_{A} = |h|x_{A} - q_{y} +\frac{\Delta_{y}}{2}$, and $z_{B} = q_{y-1}-|h|x_{B}+\frac{\Delta_{y}}{2}$. We can then express the decision rule of ML detector  as follows
\begin{align*}
     &Q\left(\frac{z_{B}-\frac{\Delta_{y}}{2} - |h|\tilde{x}}{\sqrt{\sigma^2/2}}\right) - Q\left(\frac{z_{B}+\frac{\Delta_{y}}{2} - |h|\tilde{x}}{\sqrt{\sigma^2/2}}\right)\\
     &\qquad\qquad\;\; > Q\left(\frac{z_{A}-\frac{\Delta_{y}}{2} -|h|\tilde{x}}{\sqrt{\sigma^2/2}}\right) - Q\left(\frac{z_{A}+\frac{\Delta_{y}}{2} - |h|\tilde{x}}{\sqrt{\sigma^2/2}}\right)
\end{align*}
Define the function $f(z)$ as
\begin{equation}
    \label{eq:ML_detector_rule1C}
    f(z) =  Q\left(\frac{z-\frac{\Delta_{y}}{2} - |h|\tilde{x}}{\sqrt{\sigma^2/2}}\right) - Q\left(\frac{z+\frac{\Delta_{y}}{2} - |h|\tilde{x}}{\sqrt{\sigma^2/2}}\right),
\end{equation}
\noindent which has a maximum and is symmetric at $z=|h|\tilde{x}$. Its derivative, $f'(z)$, is strictly positive (negative) for $z < |h|\tilde{x}$ ($z > |h|\tilde{x}$). Thus, $f(z)$ is higher as we go closer to $z=|h|\tilde{x}$. Equivalently,
\begingroup
\allowdisplaybreaks
\begin{align*}
f(z_{B}) > f(z_{A})  \iff \big|z_{B} -|h|\tilde{x}\big|  &< \big|z_{A} -|h|\tilde{x}\big|,
\end{align*}
which simplifies to
\begin{align}
   \bigg|q_{y-1}+\frac{\Delta_{y}}{2} -|h|x_{A}\bigg| & < \bigg|\left(q_{y}-\frac{\Delta_{y}}{2}\right)-|h|x_{B}\bigg|. 
\end{align}
\endgroup
Note that $q_{y-1}+\frac{\Delta_{y}}{2} = q_{y}-\frac{\Delta_{y}}{2} = $ middle of the ($q_{y-1},q_{y}$). If $q_{y} = +\infty$, the detection rule is 
\begin{equation}
  \label{eq:ML_detector_rule2}
    \begin{split}
    Q\left(\frac{q_{y-1} - |h|x_{A}}{\sqrt{\sigma^2/2}}\right) > &\; Q\left(\frac{q_{y-1} - |h|x_{B}}{\sqrt{\sigma^2/2}}\right)\\
    \end{split}
\end{equation}
and the inequality holds if $x_A > x_B$ because $Q(\cdot)$ is a monotic decreasing function. This can be interpreted as choosing the symbol closest to $+\infty$. Similar argument can be used for choosing the closest symbol to $-\infty$ when $q_{y-1} = -\infty$.

\section{Proof of Lemma \ref{lemma1}}\label{proof_lemma1}

We marginalize $Z\sim \text{Gamma}\left(m,\frac{\Omega}{m}\right)$ of $Q(-c+\sqrt{bz})$ for some interval $(z_{\text{lo}},z_{\text{hi}})$. Suppose we define $t_{\text{lo}} = -c + \sqrt{bz_{\text{lo}}}$, $t_{\text{hi}} = -c + \sqrt{bz_{\text{hi}}}$. $\mathcal{H}_{m,\Omega}(b,c,z_{\text{lo}},z_{\text{hi}})$ becomes
\begin{align}\label{eq:integ_exp}
    &Q(t_{\text{lo}})\frac{\Gamma\left(m,\frac{m}{\Omega}z_{\text{lo}}\right)}{\Gamma(m)}- Q(t_{\text{hi}})\frac{\Gamma\left(m,\frac{m}{\Omega}z_{\text{hi}}\right)}{\Gamma(m)}\nonumber\\
    &\qquad\qquad\qquad\qquad\qquad-\int_{t_{\text{lo}}}^{t_{\text{hi}}}\frac{e^{-\frac{t^2}{2}}\Gamma\left(m,\frac{m}{\Omega}\frac{(t + c)^2}{b}\right)}{\sqrt{2\pi}\Gamma(m)} dt
\end{align}
Since we limit our analysis to $m\in \mathbb{Z}$, We can use the following finite series representation
\begin{equation}
    \begin{split}
        \frac{\Gamma\left(m,\frac{m}{\Omega}\frac{(t + c)^2}{b}\right)}{\Gamma(m)} = \sum_{r=0}^{m-1}\frac{\left(\frac{m}{\Omega}\frac{(t + c)^2}{b}\right)^r}{r!}e^{-\frac{m}{\Omega}\frac{(t + c)^2}{b}}
    \end{split}
\end{equation}
\noindent which is based from \cite[eq. 5]{Amore:2005}. We consider two cases in our analysis of the last integral term.\\
\noindent\textbf{Case A ($c = 0$):} The integral term in (\ref{eq:integ_exp}) simplifies to
\begingroup
\allowdisplaybreaks
\begin{align}\label{eq:case_A_p1}
         & \sum_{r=0}^{m-1}\frac{\left(\frac{m}{\Omega b}\right)^r}{\sqrt{2\pi}r!}\int_{t_{\text{lo}}}^{t_{\text{hi}}}t^{2r}e^{-\frac{t^2}{2}\left(\frac{\Omega b + 2m}{\Omega b}\right)} \;dt\nonumber\\
         &=\sum_{r=0}^{m-1}\frac{\left(\frac{m}{\Omega b + 2m}\right)^r}{\sqrt{2\pi}r!}\sqrt{\frac{\Omega b}{\Omega b + 2m }}\int_{u_{\text{lo}}}^{u_{\text{hi}}}u^{2r}e^{-\frac{u^2}{2}} \;du,
\end{align}
\endgroup
\noindent where we let $u = \sqrt{\frac{\Omega b + 2m}{\Omega b}}t$, $u_{\text{lo}} = \sqrt{\frac{\Omega b + 2m}{\Omega b}}t_{\text{lo}}$, and $u_{\text{hi}} = \sqrt{\frac{\Omega b + 2m}{\Omega b}}t_{\text{hi}}$. The closed-form of the integral in the last line of (\ref{eq:case_A_p1}) is given in (\ref{eq:F_function}).
Results in (\ref{eq:integ_exp}), (\ref{eq:case_A_p1}), and (\ref{eq:F_function}) are then combined to get the expression in Lemma 1 for $c = 0$.
\\
\noindent\textbf{Case B ($c > 0$):}
Suppose we let $d = \frac{2m}{\Omega b}$, $u = \sqrt{d+1}t + \frac{dc}{\sqrt{d+1}}$, $u_{\text{lo}} = \sqrt{d+1}t_{\text{lo}} + \frac{dc}{\sqrt{d+1}}$ and $u_{\text{hi}} = \sqrt{d+1}t_{\text{hi}} + \frac{dc}{\sqrt{d+1}}$. Then, the integral in (\ref{eq:integ_exp}) simplifies to
\begingroup
\allowdisplaybreaks
\begin{align}\label{eq:caseB_p1}
        & \sum_{r=0}^{m-1}\frac{\left(\frac{d}{2}\right)^r}{\sqrt{2\pi}r!}\int_{t_{\text{lo}}}^{t_{\text{hi}}}(t + c)^{2r}e^{-\frac{1}{2}\left(t^2 + \frac{2m}{\Omega}\frac{(t+c)^2}{b}\right)} \;dt\nonumber\\
       =& \sum_{r=0}^{m-1}\frac{\left(\frac{d}{2}\right)^re^{-\frac{dc^2}{2(d+1)}}\left(\frac{c}{d+1}\right)^{2r}}{\sqrt{2\pi}r!\sqrt{d+1}}\int_{u_{\text{lo}}}^{u_{\text{hi}}}\left(\frac{\sqrt{d+1}u}{c} + 1\right)^{2r}e^{-\frac{u^2}{2}} \;du\nonumber\\
        =& \sum_{r=0}^{m-1}\sum_{l=0}^{2r}\frac{\left(\frac{d}{2}\right)^re^{-\frac{dc^2}{2(d+1)}}\left(\frac{c}{d+1}\right)^{2r}\binom{2r}{l}\left(\frac{\sqrt{d+1}}{c}\right)^l}{\sqrt{2\pi}r!\sqrt{d+1}}\int_{v_{\text{lo}}}^{u_{\text{hi}}}u^{l}e^{-\frac{u^2}{2}} \;du
\end{align}
\endgroup
The second line is obtained from the change of variable. The third line is obtained by using binomial theorem on $\left(\frac{\sqrt{d+1}u}{c} + 1\right)^{2r}$ (since $2r$ is an integer). By combining the results in (\ref{eq:integ_exp}), (\ref{eq:F_function}), and (\ref{eq:caseB_p1}), we obtain the expression in Lemma \ref{lemma1} for $c > 0$.
      

\section{Proof of Proposition \ref{prop2}}\label{proof_prop2}

Suppose $q_1^*$ is known. The region $\big[\mathcal{D}\cap\mathcal{A}\big]_{1,i}$ is fixed for all $i$.  Since $\text{supp}(Z) = [0,\infty)$ and the dependence of (\ref{eq:SEP_noiseless}) on the quantization boundaries is only through the integration bounds $\big[\mathcal{D}\cap\mathcal{A}\big]_{y,i}$, choosing a set $\{q_y'\}_{y=2}^{2^{b-1}-1}$ that simultaneously maximizes the range of $\big[\mathcal{D}\cap\mathcal{A}\big]_{y,i}$ for all $i$ and $y\in[2\mathrel{{.}\,{.}}\nobreak 2^{b-1}-1]$ also maximizes the second term in (\ref{eq:SEP_noiseless}). Consequently, $P_{\text{e},\infty}$ is minimized and $\{q_y'\}_{y=2}^{2^{b-1}-1} = \{q_y^*\}_{y=2}^{2^{b-1}-1}$. Note, however, that the range of $\big[\mathcal{D}\cap\mathcal{A}\big]_{y,i}$ for all $i$ and $y\in [2\mathrel{{.}\,{.}}\nobreak 2^{b-1}-1]$ cannot always be simultaneously maximized for general $\mathcal{X}$. In the special case where $\mathcal{X} = \mathcal{X}_{\text{g}}(\rho)$, the region $\big[\mathcal{D}\cap\mathcal{A}\big]_{y,i}$ given in (\ref{eq:DA_new_A}) can be expressed as
\begin{align*}
\frac{\max\left\{\frac{q_{y-1} + q_y}{1+\frac{1}{\rho}},q_{y-1}\right\}^2}{C^2\rho^{M-2i}}   < z <  \frac{\min\left\{\frac{q_{y-1} + q_y}{1+\rho},q_{y}\right\}^2}{C^2\rho^{M-2i}}.
\end{align*}
By inspecting the upper and lower bound of $\big[\mathcal{D}\cap\mathcal{A}\big]_{y,i}$, we see that the region is largest $\forall i \in [1\mathrel{{.}\,{.}}\nobreak \frac{M}{2}-2]$ and $\forall y \in [2\mathrel{{.}\,{.}}\nobreak 2^{b-1}-1]$ when $\frac{q_{y-1}}{q_y} = \rho$. This result also holds for (\ref{eq:DA_new_B}) and (\ref{eq:DA_new_C}). Thus, the optimal set of quantization boundaries for $\mathcal{X}_{\text{g}}(\rho)$ should satisfy (\ref{eq:q_opt_condition}).

\section{Proof of Theorem \ref{thm2}}\label{proof_thm2}

For an equidistant $M$-PAM, we have $R = \frac{M-1}{M-3}$.
We let $q_{y} = q_1 \left(R\right)^{y-1}$ to get the upper bound 
\begingroup
\allowdisplaybreaks
\begin{align}\label{eq:f_U(m,b) nonunif}
    f_{\text{U}}(m,b) =& \left(\frac{M}{4}-\frac{1}{2}\right)\left[\mathbb{P}\left(Z < \frac{ q_1^2}{\rho_1^2}\right) +  \mathbb{P}\left(Z > \frac{q^2_{1}\left(R\right)^{2^{b}-4}}{\rho_{\frac{M}{2}-2}^2}\right)\right]\nonumber\\
    =& \frac{\left(\frac{M}{4}-\frac{1}{2}\right)}{\Gamma(m)}\Bigg[\gamma\left(m,\frac{m}{\Omega}\frac{q_1^2}{\rho_1^2}\right)+\Gamma\left(m,\frac{m}{\Omega}\frac{q^2_{1}\left(R\right)^{2^b-4}}{\rho_{\frac{M}{2}-2}^2}\right)\Bigg].
\end{align}
\endgroup
The last line is obtained from the CDF of $Z\sim \mathrm{Gamma}(m,m/\Omega)$. Optimum $q_1$ is obtained by differentiating (\ref{eq:f_U(m,b) nonunif}) with respect to $q_1$ and equating the result to 0. Doing this gives us
\begin{equation}
    \begin{split}\label{eq:optimal_q1_noiseless}
        q_1^* = & \sqrt{\frac{\Omega\rho_1^2\rho_{\frac{M}{2}-2}^2}{\rho_1^2 R^{2^b-4}-\rho_{\frac{M}{2}-2}^2}\ln \left(\frac{\rho_1^2R^{2^b-4}}{\rho_{\frac{M}{2}-2}^2}\right)},
    \end{split}
\end{equation}
and the optimum value of the upper bound $f_{\text{U}}(m,b)$ is
\begin{align*}
     f_{\text{U}}(m,b) =& \frac{\left(\frac{M}{4}-\frac{1}{2}\right)}{\Gamma(m)}\Bigg[ \gamma\Bigg(m,\frac{m\rho_{\frac{M}{2}-2}^2}{\rho_1^2R^{n}-\rho_{\frac{M}{2}-2}^2}\ln \left(\frac{\rho_1^2R^n}{\rho_{\frac{M}{2}-2}^2}\right)\Bigg)\\
     &\qquad+\Gamma\Bigg(m,\frac{m\rho_1^2R^{n}}{\rho_1^2R^{n}-\rho_{\frac{M}{2}-2}^2}\ln \left(\frac{\rho_1^2R^n}{\rho_{\frac{M}{2}-2}^2}\right)\Bigg)\Bigg],
\end{align*}
where we let $n = 2^b-4$. Suppose there is sufficiently large $b$ such that
\begin{align*}
    R^{2^{b}-4} >& \left(\frac{\rho_{\frac{M}{2}-2}^2}{\rho_1^2}\right)^2,
\end{align*}
or equivalently,
\begin{align}
\label{eq:sufficient_b}
    b > \log_2\Bigg[\frac{\log_2\left(\rho_{\frac{M}{2}-2}^2\right) - \log_2\left(\rho_1^2\right)}{\log_2\left(R\right)} +2\Bigg]+1.
\end{align}
Then, $f_{\text{U}}(m,b)$ is asymptotically equivalent to
\begingroup
\allowdisplaybreaks
\begin{align}
  f_{\text{U}}(m,b)
    \;&\sim_{m,n}^{\infty,\infty}\;   \frac{\left(\frac{M}{4}-\frac{1}{2}\right)}{\Gamma(m)}\Bigg[\gamma\Bigg(m,m\frac{\rho_{\frac{M}{2}-2}^2}{\rho_1^2R^{n}}\ln \left(\frac{\rho_1^2R^n}{\rho_{\frac{M}{2}-2}^2}\right)\Bigg)\nonumber\\
    &\qquad\qquad\qquad+\Gamma\left(m,m\ln \left[\frac{\rho_1^2R^n}{\rho_{\frac{M}{2}-2}^2}\right]\right)\Bigg]\nonumber\\
    \;&\sim_{m,n}^{\infty,\infty}\;\left(\frac{M}{4}-\frac{1}{2}\right)\frac{m^{-\frac{1}{2}}e^{m-1}}{\sqrt{2\pi}}\left\{\frac{\rho_{\frac{M}{2}-2}^2}{\rho_1^2R^n}\right\}^m\nonumber\\
    &\qquad\qquad\times\left\{\ln\left[\frac{\rho_1^2R^n}{\rho_{\frac{M}{2}-2}^2}\right]\right\}^{m-1}\left\{\ln\left[\frac{\rho_1^2R^n}{\rho_{\frac{M}{2}-2}^2}\right] + 1 \right\} \nonumber\\
    &\;=\;O\left(\frac{2^{b-1}m^{-\frac{1}{2}}2^m}{R^{2^bm}}\left\{\ln\left[\frac{1}{R}\right]\right\}^m\right)\nonumber\\
    &\;=\;O\left(2^{-\left[(2^b-1)m-b\right]}\right)
\end{align}
\endgroup
The second line comes from the limiting behavior of the upper and lower incomplete Gamma functions \cite{Jameson:2016} (i.e. $\Gamma(a,x) \sim_x^\infty x^{a-1}e^{-x}$ and $\gamma(a,x) \sim_x^0 \frac{x^{a}}{a}$) and Stirling's approximation of the factorial function. The third line is obtained by representing the asymptotic behavior in terms of Big-O notation and noting that $n \sim_b^\infty 2^b$. It is straightforward to show that $f_{\text{L}}(m,b)$ is asymptotically equivalent to $f_{\text{U}}(m,b)$ up to a scaling factor. Thus, $P^{*M}_{\text{e},\infty}(m,b)$ also follows the same asymptotic behavior.

\section{Proof of Theorem \ref{thm3}}\label{proof_thm3}

We can think of an SEP-optimal uniform quantizer as a special case of SEP-optimal quantization which has a constraint $\Delta_q = q_{y}-q_{y-1}\;\;\forall y \in [1\mathrel{{.}\,{.}}\nobreak 2^{b-1}-1]$. We use $q_y = y\Delta_q$ for the $y$-th quantization boundary. The proof of Theorem \ref{thm3} follows a similar approach to that of Theorem \ref{thm2}. Using uniform quantization and the CDF of $Z\sim \mathrm{Gamma}(m,m/\Omega)$, (\ref{eq:f_U(m,b)}) specializes to
 \begin{align}\label{eq:f_U(m,b) unif}
        f_{\text{U}}(m,b) 
        =& \left(\frac{M}{4}-\frac{1}{2}\right)\Bigg[\frac{\gamma\left(m,\frac{m}{\Omega}\frac{\Delta_q^2}{\rho_1^2}\right)+\Gamma\left(m,\frac{m}{\Omega}\frac{(2^{b-1}-1)^2\Delta^2_{q}}{\rho_{\frac{M}{2}-2}^{2}}\right)}{\Gamma(m)}\Bigg].
    \end{align}
    The optimal quantization step, $\Delta_q^*$, obtained by differentiating (\ref{eq:f_U(m,b) unif}) and equating the result to 0, is
\begin{equation}\label{eq:delta_opt_standard_unif}
    \Delta_q^* =\sqrt{ \frac{\Omega \rho_1^2\rho_{\frac{M}{2}-2}^2}{\rho_1^2(2^{b-1}-1)^2-\rho_{\frac{M}{2}-2}^2}\ln\left[\left(\frac{\rho_1(2^{b-1}-1)}{\rho_{\frac{M}{2}-2}}\right)^2\right]}.
\end{equation}
 If there is sufficiently large $b$ such that 
\begin{align}
    (2^{b-1}-1)^{2} > \left(\frac{\rho_{\frac{M}{2}-2}}{\rho_1}\right)^2\;\Rightarrow\; b > \log_2\left(\frac{\rho_{\frac{M}{2}-2}}{\rho_1}+1\right) + 1,
\end{align}
then the asymptotic behavior of (\ref{eq:f_U(m,b) unif}) for increasing $m$ and $b$ is
\begingroup
\allowdisplaybreaks
\begin{align}
    f_{\text{U}}(m,b) \;&\sim_{m,n}^{\infty,\infty}\;  \frac{\left(\frac{M}{4}-\frac{1}{2}\right)}{\Gamma(m)}\Bigg[\gamma\left(m,2m\frac{\rho_{\frac{M}{2}-2}^2}{n}\ln\left(\frac{\rho_1n}{\rho_{\frac{M}{2}-2}}\right)\right)\nonumber\\
    &\qquad\qquad\qquad+\Gamma\left(m,2m\ln\left(\frac{\rho_1n}{\rho_{\frac{M}{2}-2}}\right)\right)\Bigg]\nonumber\\
    \;&\sim_{m,n}^{\infty,\infty}\; \left(\frac{M}{4}-\frac{1}{2}\right)\frac{m^{-\frac{1}{2}}\{2e\}^{m-1}}{\sqrt{2\pi}}\left\{\frac{\rho_{\frac{M}{2}-2}}{n}\right\}^{2m}\nonumber\\
    &\qquad\qquad\times\left\{\ln\frac{\rho_1n}{\rho_{\frac{M}{2}-2}}\right\}^{m-1}\left\{\ln\frac{\rho_1 n}{\rho_{\frac{M}{2}-2}}+\frac{1}{\rho_1^{2m}}\right\}\nonumber\\
    &\;=\;O\left(2^{-2bm}\right)
\end{align}   
\endgroup
where we let $n = 2^{b-1}-1$ and used $\Delta_q^*$. Second line comes from the limiting behavior of the upper and lower incomplete Gamma functions \cite{Jameson:2016} and Stirling's approximation of the factorial function. The third line is obtained by representing the asymptotic behavior in terms of Big-O notation and that $n \sim_b^\infty 2^b$. It is straightforward to show that $f_{\text{L}}(m,b)$ is asymptotically equivalent to $f_{\text{U}}(m,b)$ up to a scaling factor. Thus, $P^{*M}_{\text{e},\infty}(m,b)$ also follows the same asymptotic behavior.

\section{Proof of Theorem \ref{thm4}}\label{proof_thm4}

 Suppose we use the $M$-PAM constellation in (\ref{eq:M-PAM constellation error floor}). By Corollary \ref{corollary 1}, we proved that SEP can be reduced to zero if $2^b > M-2$. We start the proof with Lemma \ref{lemma6}.
        \begingroup
        \allowdisplaybreaks
        \begin{align}\label{eq:Pe_highSNR_approx}
            P^M_e\left(\frac{1}{\sigma^2}\right) \doteq& \left[\frac{\sigma^2}{\rho_0^2}\right]^{m}+ \underbrace{\sum_{i = 0}^{\frac{M}{2}-1}\sum_{n=0,n\ne i}^{\frac{M}{2}-1}\mathbb{P}(\hat{x} = +\rho_n|x = +\rho_i)}_{\text{primarily affected by fading as $\sigma^2\rightarrow 0$}}\nonumber\\
            \doteq& \left[\frac{\sigma^2}{C^2\rho^M}\right]^{m} + \kappa\frac{\gamma\left(m,\frac{mq_1^2}{C^2\rho^{M-2}}\right)+\Gamma\left(m,\frac{mq^2_{1}}{C^2\rho^{2^b}}\right)}{\Gamma(m)}\nonumber\\
            \doteq& \left[\frac{\sigma^2}{C^2\rho^M}\right]^{m} + \frac{\gamma\left(m,\frac{mq_1^2}{C^2\rho^{M-2}}\right)+\Gamma\left(m,\frac{mq^2_{1}}{C^2\rho^{2^b}}\right)}{\Gamma(m)},
    \end{align}
    \endgroup
     where we note in the underbraces of the first line that the last two terms are primarily affected by fading for arbitrarily large SNR. Equation (\ref{eq:errorfloor M-PAM shaping}) is used for the second summation term with some scaling factor $\kappa\in \left[\frac{2}{M}, \frac{M}{4}-\frac{1}{2}\right]$. Using $\kappa = \frac{2}{M}$ and $\kappa = \frac{M}{4}-\frac{1}{2}$ gives the lower bound and upper bound of (\ref{eq:errorfloor M-PAM shaping}), respectively. The last line is obtained by using Lemma \ref{lemma5}.ii. $q_1^*(\rho)$ is obtained by differentiating equation (\ref{eq:Pe_highSNR_approx}) with respect to $q_1$ and equating the result to 0.
    \begin{equation}\label{eq:q_1_opt_nonunif}
        \begin{split}
            q_1^*\left(\rho\right) = & \sqrt{\frac{\rho^{2^b}C^2(2^b-M+2)}{1-\rho^{2^b-M+2}}\ln\left(\frac{1}{\rho}\right)}
        \end{split}
    \end{equation}
    By Corollary \ref{corollary 1}, $\rho$ should be arbitrarily small at high SNR regime to make the SEP also arbitrarily small. Using $q_1^*$ and $n = 2^{b}-M+2$, the second term of ($\ref{eq:Pe_highSNR_approx}$) is asymptotically equivalent to
\begingroup
\allowdisplaybreaks
 \begin{align}\label{eq:Pe_asymptotic}
     \frac{\gamma\left(m,\frac{mq_1^{*2}}{C^2\rho^{M-2}}\right)+\Gamma\left(m,\frac{mq^{*2}_{1}}{C^2\rho^{2^b}}\right)}{\Gamma(m)}
      \;\sim_{\rho}^0&\; \frac{\left[mn\right]^{m-1}\left[\ln\left(\frac{1}{\rho}\right)\right]^m\rho^{mn}}{\Gamma(m)}
\end{align}
\endgroup
By substituting (\ref{eq:Pe_asymptotic}) to the second term of the right-hand side of (\ref{eq:Pe_highSNR_approx}) and applying Lemma \ref{lemma5}.ii, we get
\begingroup
\allowdisplaybreaks
\begin{align}
    P^M_e\left(\frac{1}{\sigma^2}\right)
     \doteq&\left[\frac{\sigma^2}{C^2\rho^M}\right]^{m} + \left[\ln\left(\frac{1}{\rho}\right)\right]^{m}\rho^{mn}.
    \end{align}
    \endgroup
This substitution is valid by Corollary \ref{corollary 1}. Note that $\ln(x) \geq 1,\; \forall x \geq e$ and any polynomial $P(x)$ with degree $p > 0$ grows faster than logarithm function for arbitrarily large $x$. Thus, the following inequality holds:
\begin{equation}\label{eq:ln_asymptotic}
     \rho^{a} \leq\left[\ln\left(\frac{1}{\rho}\right)\right]^{m}\rho^{a} \leq \; \rho^{a - m\epsilon}
\end{equation}
for some $a > 0$ and arbitrarily small $\rho$,$\epsilon$. By squeeze theorem, the expression simplifies to
\begingroup
\allowdisplaybreaks
\begin{align}\label{eq:PE_highSNR_approx_simplify}
     P^M_e\left(\frac{1}{\sigma^2}\right)\doteq&\left[\frac{\sigma^2}{C^2\rho^M}\right]^{m} +\rho^{mn}\nonumber\\
     \doteq& \left[\frac{\sigma^2}{\rho^{M-2}}\right]^{m} +\rho^{mn}\qquad\left(\text{since $C^2 \sim_\rho^0 \frac{1}{\rho^2}$}\right)\nonumber\\
     \doteq& \left[\frac{1}{\sigma^2}\right]^{-\frac{mn}{n+(M-2)}}\qquad\qquad\text{(by Lemma \ref{lemma7}).}
\end{align}
\endgroup
From the definition of decay exponent, the DVO of (\ref{eq:PE_highSNR_approx_simplify}) is $\frac{mn}{n+(M-2)} = m\frac{2^{b}-M+2}{2^{b}}$. The optimality of this DVO comes from choosing $\rho^*$ depending on $\sigma^2$ and optimizing $q_1^*$ based on $\rho$. 

\section{Proof of Theorem \ref{thm5}}\label{proof_thm5}

Suppose we use the constellation in (\ref{eq:M-PAM constellation error floor}) for $M = 4$. We start the proof with Lemma \ref{lemma6}.
        \begingroup
        \allowdisplaybreaks
        \begin{align}\label{eq:Pe_highSNR_approx_unif}
            P^4_e\left(\frac{1}{\sigma^2}\right) \doteq& \left[\frac{\sigma^2}{\rho_0^2}\right]^{m}+ \underbrace{\sum_{i = 0}^{\frac{M}{2}-1}\sum_{n=0,n\ne i}^{\frac{M}{2}-1}\mathbb{P}(\hat{x} = +\rho_n|x = +\rho_i)}_{\text{primarily affected by fading as $\sigma^2\rightarrow 0$}}\nonumber\\
            \doteq& \left[\frac{\sigma^2}{C^2\rho^4}\right]^{m} + \Bigg[\frac{\gamma\left(m,m\frac{\Delta_q^2}{C^2\rho^{2}}\right)}{\Gamma(m)}\nonumber\\
            &\qquad+\frac{\Gamma\left(m,m\frac{(2^{b-1}-1)^2\Delta^2_{q}}{C^2\rho^{4}}\right)}{\Gamma(m)}\Bigg],
    \end{align}
    \endgroup
where we note in the underbraces of the first line that the last two terms are primarily affected by fading for arbitrarily large SNR. Equation (\ref{eq:errorfloor M-PAM shaping}) is used for the second summation term but the coefficient is dropped in last line due to Lemma \ref{lemma5}.ii. $\Delta_q^*(\rho)$ is obtained by differentiating equation (\ref{eq:Pe_highSNR_approx_unif}) with respect to $\Delta_q$ and equating the result to 0.
\begin{equation}\label{eq:delta_opt_unif}
    \Delta_q^* = \sqrt{\frac{2\rho^4C^2}{(2^{b-1}-1)^2-\rho^2}\ln\left[\frac{2^{b-1}-1}{\rho}\right]}.
\end{equation}
By Corollary \ref{corollary 1},  $\rho$ should be arbitrarily small at high SNR regime to make the SEP also arbitrarily small. Using $\Delta_q^*$ and $n = 2^{b-1}-1$, the second term of ($\ref{eq:Pe_highSNR_approx_unif}$) is asymptotically equivalent to
\begingroup
\allowdisplaybreaks
 \begin{align}\label{eq:Pe_asymptotic_unif}
    &\frac{\gamma\left(m,m\frac{\Delta_q^{*2}}{C^2\rho^{2}}\right)+\Gamma\left(m,m\frac{(2^{b-1}-1)^2\Delta^{*2}_{q}}{C^2\rho^{4}}\right)}{\Gamma(m)}\nonumber\\
      &\qquad\qquad\qquad\quad\sim_{\rho}^0\; \frac{\{2m\}^{m-1}\left[\ln n + \ln\frac{1}{\rho}\right]^{m}}{\Gamma(m)}\left[\frac{\rho}{n}\right]^{2m}.
\end{align}
\endgroup
By substituting (\ref{eq:Pe_asymptotic_unif}) to the second term of (\ref{eq:Pe_highSNR_approx_unif}) and applying Lemma \ref{lemma5}.i and \ref{lemma5}.ii, we get
\begin{align}
    P^4_e\left(\frac{1}{\sigma^2}\right)\doteq&\left[\frac{\sigma^2}{C^2\rho^4}\right]^{m} + \left[\ln\left(\frac{1}{\rho}\right)\right]^{m}\rho^{2m}.
\end{align}
This substitution is valid by Corollary \ref{corollary 1}. Using (\ref{eq:ln_asymptotic}) and the fact that $C^2 \sim_\rho^0 \frac{1}{\rho^2}$, the expression simplifies to
\begin{align}\label{eq:PE_highSNR_approx_simplify_unif}
     P^4_e\left(\frac{1}{\sigma^2}\right)
     \doteq& \left[\frac{\sigma^2}{\rho^{2}}\right]^{m} +\rho^{2m}\nonumber\\
     \doteq& \left[\frac{1}{\sigma^2}\right]^{-\frac{m}{2}}\quad\left(\text{by Lemma \ref{lemma7}}\right).
\end{align}
From the definition of decay exponent, the DVO of (\ref{eq:PE_highSNR_approx_simplify_unif}) is $\frac{m}{2}$ when $\rho^*$ is used. The optimality of this DVO comes from choosing $\rho^*$ depending on $\sigma^2$ and optimizing $q_1^*$ based on $\rho$. 

\end{appendices}


\ifCLASSOPTIONcaptionsoff
  \newpage
\fi

\bibliographystyle{ieeetr}
\bibliography{references}
\vskip 0pt plus -1fil
\begin{IEEEbiography}[{\includegraphics[width=1in,height=1.25in,clip,keepaspectratio]{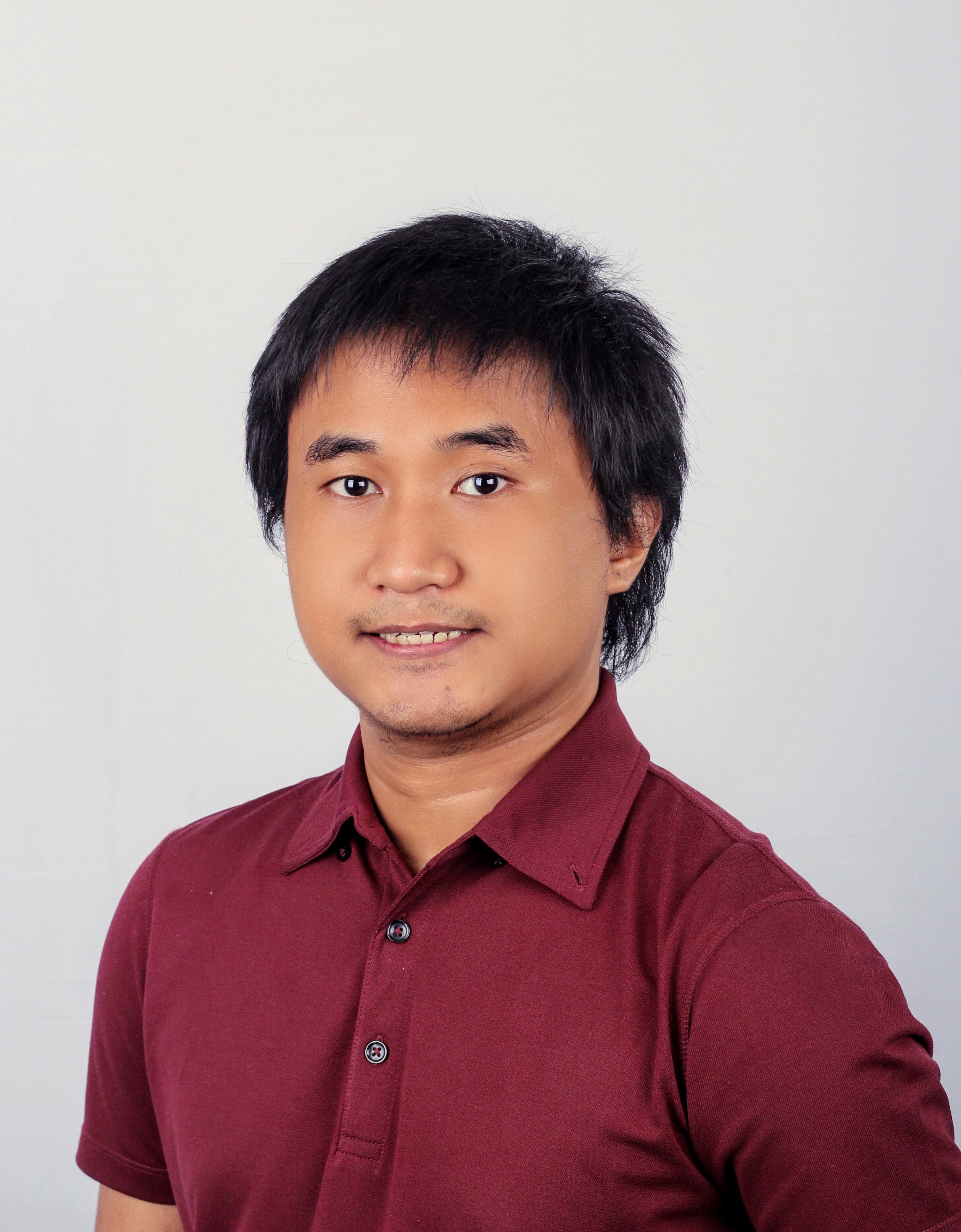}}]{Neil Irwin Bernardo} received his B.S. degree in Electronics and Communications Engineering from the University of the Philippines Diliman in 2014 and his M.S. degree in Electrical Engineering from the same university in 2016. He has been a faculty member of the University of the Philippines Diliman since 2014, and is currently on study leave to pursue a Ph.D. degree in Engineering at the University of Melbourne, Australia. His research interests include wireless communications, signal processing, and information theory.
\end{IEEEbiography}
\vskip 0pt plus -1fil
\begin{IEEEbiography}[{\includegraphics[width=1in,height=1.25in,clip,keepaspectratio]{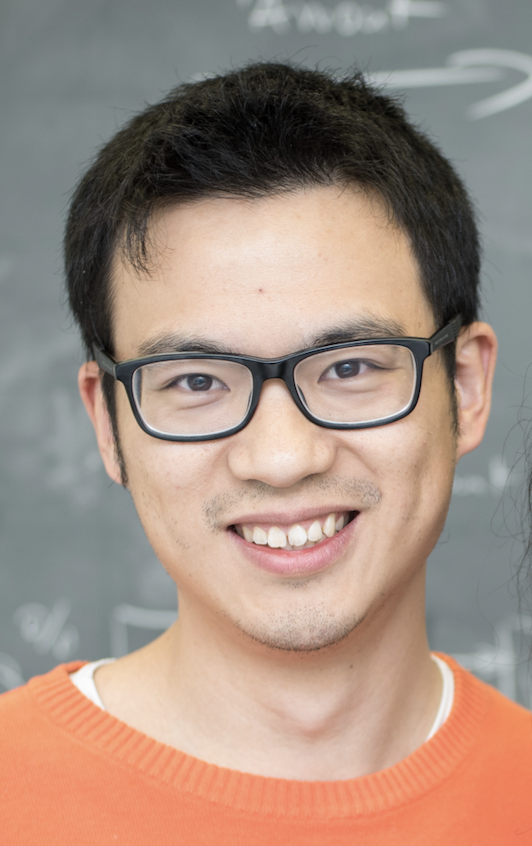}}]{Jingge Zhu} received the B.S. degree and M.S. degree in electrical engineering from Shanghai Jiao Tong University, Shanghai, China, in 2008 and 2011, respectively, the Dipl.-Ing. degree in technische Informatik from Technische Universit\"{a}t Berlin, Berlin, Germany in 2011 and the Doctorat \`{e}s Sciences degree from the Ecole Polytechnique F\'{e}d\'{e}rale (EPFL), Lausanne, Switzerland, in 2016. He was a post-doctoral researcher at the University of California, Berkeley from 2016 to 2018.  He is now a lecturer at the University of Melbourne, Australia. His research interests include information theory with applications in communication systems and machine learning. 

Dr. Zhu received the Discovery Early Career Research Award (DECRA) from the Australian Research Council in 2021, the IEEE Heinrich Hertz Award for Best Communications Letters in 2013, the Early Postdoc. Mobility Fellowship from Swiss National Science Foundation in 2015, and the Chinese Government Award for Outstanding Students Abroad in 2016.
\end{IEEEbiography}
\vskip 0pt plus -1fil
\begin{IEEEbiography}[{\includegraphics[width=1in,height=1.25in,clip,keepaspectratio]{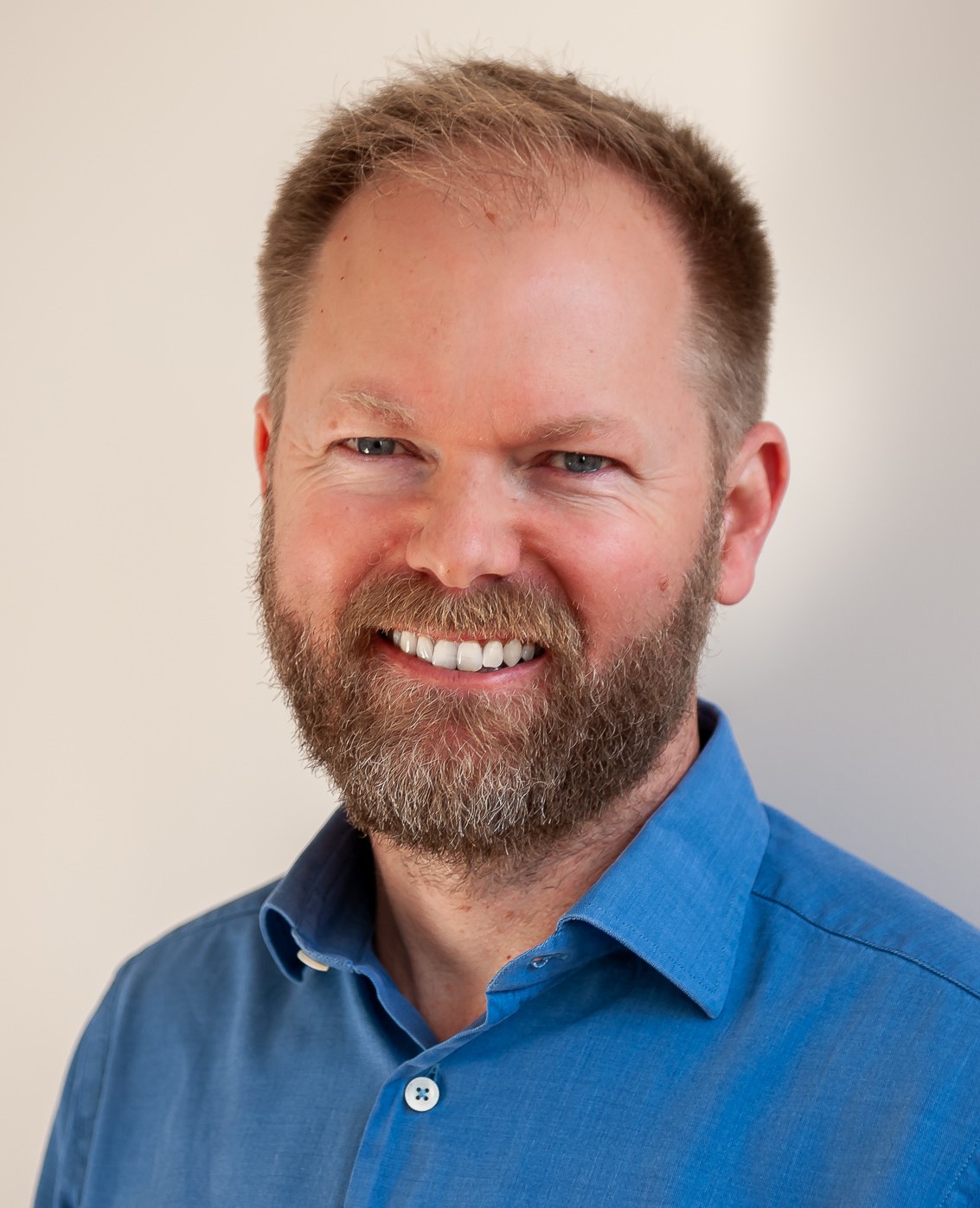}}]{Jamie Evans} was born in Newcastle, Australia, in 1970. He received the B.S. degree in physics and the B.E. degree in computer engineering from the University of Newcastle, in 1992 and 1993, respectively, where he received the University Medal upon graduation. He received the M.S. and the Ph.D. degrees from the University of Melbourne, Australia, in 1996 and 1998, respectively, both in electrical engineering, and was awarded the Chancellor's Prize for excellence for his Ph.D. thesis. From March 1998 to June 1999, he was a Visiting Researcher in the Department of Electrical Engineering and Computer Science, University of California, Berkeley. Since returning to Australia in July 1999 he has held academic positions at the University of Sydney, the University of Melbourne and Monash University. He is currently a Professor of Electrical and Electronic Engineering and Pro Vice-Chancellor (Education) at the University of Melbourne. His research interests are in communications theory, information theory, and statistical signal processing with a focus on wireless communications networks.
\end{IEEEbiography}
\end{document}